\documentclass[sts]{imsart}

\RequirePackage{amsthm,amsmath,amsfonts,amssymb}
\RequirePackage[numbers]{natbib}
\RequirePackage{graphicx}%% uncomment this for including figures
\usepackage{stmaryrd}

\startlocaldefs
\endlocaldefs

\begin{document}

\begin{frontmatter}
%%%%%%%%%%%%%%%%%%%%%%%%%%%%%%%%%%%%%%%%%%%%%%
%%                                          %%
%% Enter the title of your article here     %%
%%                                          %%
%%%%%%%%%%%%%%%%%%%%%%%%%%%%%%%%%%%%%%%%%%%%%%
\title{Topological Data Analysis for Multivariate
Time Series Data}
%\title{A sample article title with some additional note\thanksref{T1}}
\runtitle{Topological Data Analysis for Multivariate
Time Series Data}
%\thankstext{T1}{A sample of additional note to the title.}

\begin{aug}
\author[A]{\fnms{Anass} \snm{El Yaagoubi Bourakna}\ead[label=e1]{anass.bourakna@kaust.edu.sa}},
\author[B]{\fnms{Moo} \snm{K. Chung}\ead[label=e2]{mkchung@wisc.edu}},
and
\author[C]{\fnms{Hernando} \snm{Ombao}\ead[label=e3]{hernando.ombao@kaust.edu.sa}}
%%%%%%%%%%%%%%%%%%%%%%%%%%%%%%%%%%%%%%%%%%%%%%
%% Addresses                                %%
%%%%%%%%%%%%%%%%%%%%%%%%%%%%%%%%%%%%%%%%%%%%%%
\address[A]{KAUST, \printead{e1}.}
\address[B]{University of Wisconsin-Madison, \printead{e2}.}
\address[C]{KAUST, \printead{e3}.}
\end{aug}

\begin{abstract}
    Over the last two decades, topological data analysis (TDA) has emerged as a very powerful data analytic approach which can deal with various data modalities of varying complexities. One of the most commonly used tools in TDA is persistent homology (PH) which can extract topological properties from data at various scales. Our aim in this article is to introduce TDA concepts to a statistical audience and provide an approach to analyze multivariate time series data. The application focus will be on multivariate brain signals and brain connectivity networks.    Finally, the paper concludes with an overview of some open problems and potential application of TDA to modeling directionality in a brain network as well as the casting of TDA in the context of mixed effects models to capture variations in the topological properties of data collected from multiple subjects.
\end{abstract}


\end{frontmatter}

%%%%%%%%%%%%%%%%%%%%%%%%%%%%%%%%%%%%%%%%%%%%%%
%%%% Main text entry area:

\newpage
\section{Introduction}
\label{sec:introduction}

The field of topology can be traced back more than two centuries ago, starting from the work of Leonhard Euler on the famous Königsberg bridge problem, which consists in finding a walk through the city that would cross each of those bridges once and only once \citep{TOPOLOGY_EULER}. As mentioned in \cite{TOPOLOGY_HISTORY}, for over a century later, the field of topology was enriched by the contributions of numerous renowned mathematicians, such as Enrico Betti, Camille Jordan, Johann Benedict Listing, Bernhard Riemann, Felix Hausdorff (much later) and many others. By the turn of the century, Henri Poincaré had developed the concepts of homotopy and homology, thus starting the new field of algebraic topology. The field of topology witnessed major advances and theoretical breakthroughs throughout the twentieth century, becoming one of the most important fields of mathematics, but without practical applications. 

Despite the major theoretical development throughout the twentieth century, the application aspect did not really take off until much later. Indeed, was only the beginning of this century that topology found its prosperous way to the applications arena under the coinage of topological data analysis. Topological data analysis (or TDA) has witnessed many important advances over the last twenty years, it aims to unravel and provide insights about the "shape" of the data, following the central TDA dogma: data has shape, shape has meaning and meaning drives value. This is done by analyzing the persistence homology using a persistence diagram or barcode. The reader is refered to \cite{TDA_EDELSBRUNNER} for an introduction to the notion of persistence, \cite{EDELSBRUNNER_HARER} and \cite{BARCODES} for a survey on persistent homology and barcodes, and \cite{TDA_GUNNAR} for a review of TDA in general.

Several tools have been developed under the TDA framework to analyze all sorts of data. These tools have been applied to various scientific fields including: Biology  \cite{TDA_BIOLOGY}, finance \cite{TDA_FINANCIAL_TSA} and brain signals \cite{TDA_BRAIN} and \cite{TDA_BRAIN_ARTERY}. These tools aim to guide the practitioner to understand the geometrical features present in high dimensional data, which are not always directly accessible using other classical techniques. Even if such features could be observed upon examination using classical graph theoretical methods, these would not have been found automatically if the topological methods had not first detected them. As a summary, TDA provides a set of techniques for dimensionality reduction without loss in the topological information contained in the data.

The main tools in TDA are Morse and Vietoris-Rips filtrations. These techniques have been extensively used in various applications. For instance, Morse filtrations have been used to study patterns in imaging data such as in \cite{TDA_MORSE_CORTICAL_DATA} and \cite{TDA_MORSE_EEG}, or the geometry of random fields in general as in \cite{TDA_MORSE_RANDOM_FIELDS}. However, Vietoris-Rips filtrations (through persistence homology) have been extensively used to study all sorts of point cloud data sets. This includes time series data and its transformations using various embedding techniques. To better understand persistence homology, a few concepts such as simplicial complexes and their filtrations will be introduced.

The overall objective of this paper is to introduce some of the key TDA concepts to a statistical audience and provide an overview of the potential applications of these concepts to multivariate time series data. Thus, we briefly introduce TDA and persistent homology, and focus on various approaches that provide a powerful way to analyze multivariate time series data. In Section 2 we review some background material for TDA: Morse filtration, the persistence homology and time delay embeddings of univariate time series. In Section 3, we present TDA for dependence networks of multivariate time series. In Section 4, we propose a testing framework for identifying topological group differences based on permutation tests. Finally, in Section 5 we present various open problems of interest in brain dependence studies.

% \footnote{The last paragraph can be improved. Many sentences are repeating.}

\section{Background material for TDA}

Understanding topological data analysis requires familiarization with a few key concepts. When reading Topological Data Analysis (TDA) literature, it is common to find terminology such as data points, distance between points, or whether the following TDA summary is stable or unstable. Such phrases often use vocabulary familiar to the statistician, however the meaning may differ greatly, making it difficult for the reader to understand. Therefore, we invite the reader to ask themselves the following questions:
\begin{itemize}
    \item Meaning of data: What constitutes data?
    \item Meaning of distance: How can we define a meaningful distance between data points?
    \item Notion of stability: Is this given TDA summary stable? This is addressed through stability theorems.
\end{itemize}
These questions will be addressed (directly or indirectly) in this section as well as the coming sections. First, we start by investigating the use of TDA with univariate time series data, then we will consider multivariate time series data such as brain electroencephalogram signals recorded from many electrodes on the scalp. Furthermore, we will examine dependence-based distance functions that measures the degree of association between time series components (e.g., between pairs of electrodes). Finally, we demonstrate TDA on real-world applications with dependence networks. 

There is no doubt that data encompasses a wide range of concepts. In the mind of a geneticist, data means something specific (e.g., sequence of nucleotides), however, data means something else for the neuroscientist who works with brain signals (e.g., Electroencephalogram (EEG), Local Field Potential (LFP), functional Magnetic Resonance Imaging (fMRI) etc.). In general data can be represented in various forms; for example, images, functions, time series, counts, random fields, bandpass-filtered signals, Fourier transforms, localized transforms (e.g., wavelets and SLEX) etc. In other words, every type of data may require a different statistical approach. For example, a cloud of points in Euclidean space might require one statistical approach, while a time series of counts might require another.

Similarly, the notion of distance could potentially vary across data modalities. Indeed, the notion of distance is intrinsically linked to the nature of the data. Distance means some measure of proximity between data points present in some space of reference. For instance, it is meaningless to use a Euclidean distance when dealing with categorical data as opposed to continuous data, e.g., gender data or a count time series vs a cloud of points in $\mathcal{R}^2$. 

In general, the notion of distance aims to capture some notion of similarity. When the data naturally originates from a meaningful metric space, one can use the inherited distance metric from the space of reference. Otherwise, a more suitable metric should be used to capture the information of interest. For example, if we are interested in studying the brain networks originating from brain imaging techniques, a meaningful distance metric could be based on the notion of dependence. Sometimes, the goal in a study is to examine the extent of synchrony between regions in a brain network and how that synchrony may be disrupted due to a stimulus or a shock. Often it would be more informative to study the potential cross-interactions between oscillatory activities at different channels and how an increased amplitude in one channel may excite (increase the amplitude) or inhibit (decrease the amplitude) of another channel, see \cite{SPECTRAL_DEPENDENCE}. Simple correlation-based distance measures have become commonplace in many applications due to their simplicity in computation and interpretation. However, these approaches can only examine linear associations and may not be appropriate for brain signals, where frequency-specific cross-interactions may be more informative.

Stability is a tricky concept mainly because it has many facets, especially in the context of TDA. While the origins of TDA have been in mathematics, it is now more often applied and further developed by statisticians who may have different motivations and applications in mind. In the mind of a mathematician, the stability of some transformation ($g$) means robustness to small deformations in the input and is usually captured by inequalities such as:
\begin{align}
    d_{F}\big(g(X), g(Y)\big) \leq d_{E}\big(X, Y\big), \text{ where } g : E \to F \label{eq:stability_inequality}
\end{align}
where $Y$ is a smooth transformation of $X$ and $g$ is the transformation for which we want to prove the stability. Where $d_E$ could be the Euclidean distance or the Hausdorff distance between $X$ and $Y$ and $d_F$ could be the Frobenius norm ($||.||_F$) if the function $g$ returns a matrix or it could be the Wasserstein or Bottleneck distance (as defined in Equations \ref{Eq:wasserstein_distance} and \ref{Eq:bottleneck_distance}) if $g$ return a set of points.

\begin{align}
    W_p(A, B) &= \underset{\gamma:A \twoheadrightarrow B}{\inf} \hspace{5mm} \Bigg( \sum_{x \in A} \big|\big| x-\gamma(x) \big|\big|_\infty^p \Bigg)^\frac{1}{p} \label{Eq:wasserstein_distance},\\
    W_\infty(A, B) &= \underset{\gamma:A \twoheadrightarrow B}{\inf} \hspace{5mm} \underset{x \in A}{sup} \big|\big| x-\gamma(x) \big|\big|_\infty \label{Eq:bottleneck_distance},
\end{align}
where $\gamma$ represents a bijection between the sets $A$ and $B$. 

For statisticians, stability has a completely different meaning. Indeed, statistical reasoning considers stability based on the robustness of a conclusion, e.g., the result of a statistical test or inference, against perturbations in the original data due to random noise or any departure from initial model assumptions. When the mathematician considers small/smooth perturbations in the input, the statistician however, considers adding random noise with small standard deviation or a set of outliers to the data. Furthermore, the statistician might also consider the perturbation in terms of a change in the distribution of the data, e.g., if the distribution of the data has a heavier tail (for example student's t-distribution instead of a normal distribution), this might result in the presence of many unexpected outliers.

\begin{figure}
    \centering
    \includegraphics[width=.8\linewidth]{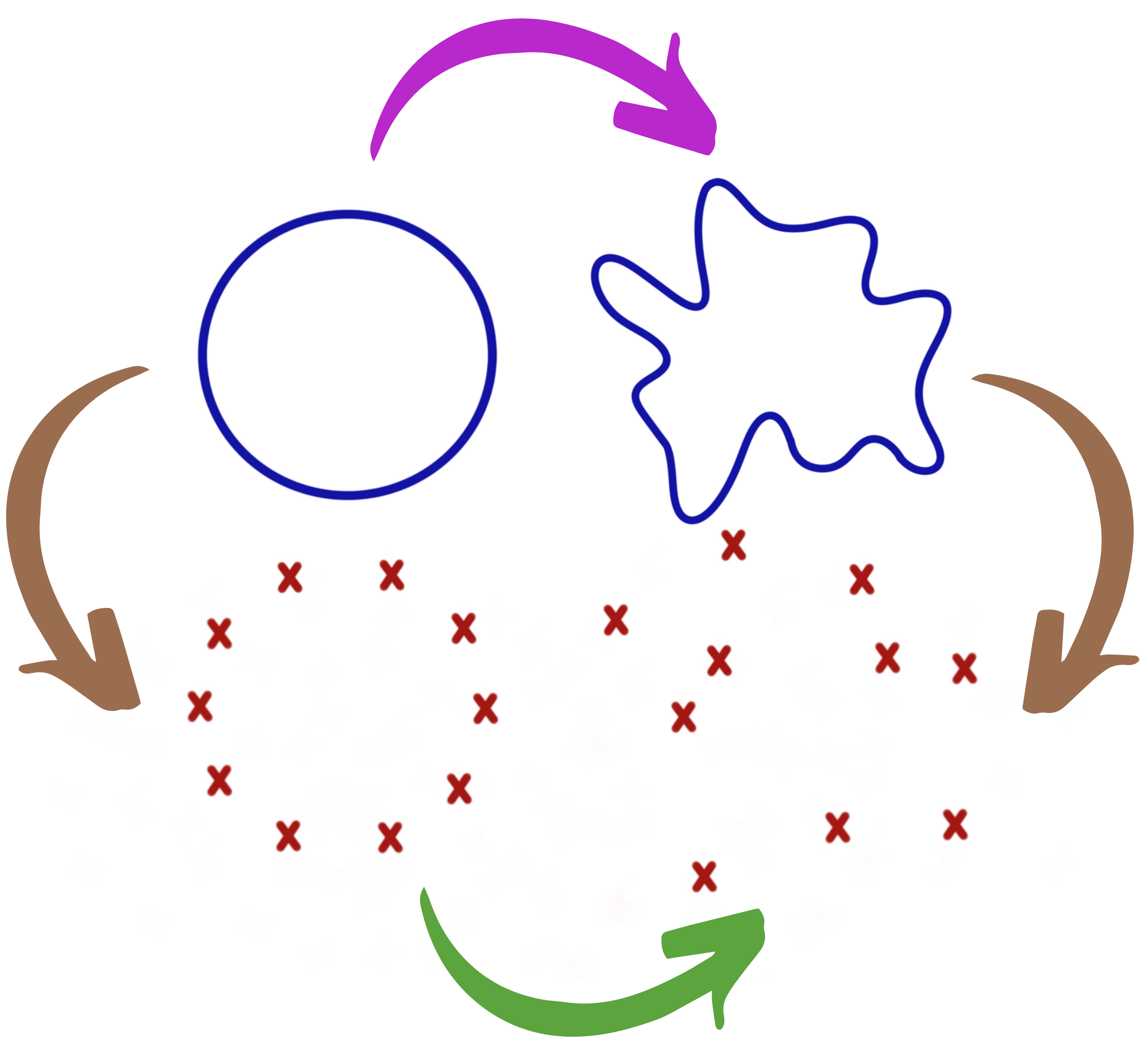}
    \caption{Perturbation of the data from the perspectives of statistics and topology. Original topological structure is in the upper-left corner (circle). Observed cloud of points in the bottom-right corner. The mathematician first considers a perturbation of the manifold (PURPLE arrow) then a sampling step (BROWN right arrow), whereas the statistician first considers a sampling step (BROWN left arrow) then the addition of noise (GREEN arrow).}
    \label{fig:manifold_sampling}
\end{figure}

Both perspectives can lead to the same result in some cases. For instance, if data is sampled from a manifold (later in Figure \ref{fig:time_series_embedding_illustration_3} we provide an illustration of this abstract notion of sampling time series components from an underlying dependence manifold) and small sampling noise is considered, it is typically similar to smoothly deforming the manifold and resampling from the new manifold as seen from Figure \ref{fig:manifold_sampling}. However, in this context, adding an outlier to the data may alter the topology of the object completely. In this case, having a stability theorem is of little practical importance since it is too restrictive. Indeed, adding an outlier completely alters the input. Hence, the inequality in Equation \ref{eq:stability_inequality} still holds. However, it is not always tight enough to provide meaningful conclusions. More details regarding the persistence diagram can be found in \cite{PD_STABILITY}.

\begin{figure}
    \centering
    \includegraphics[width=\linewidth]{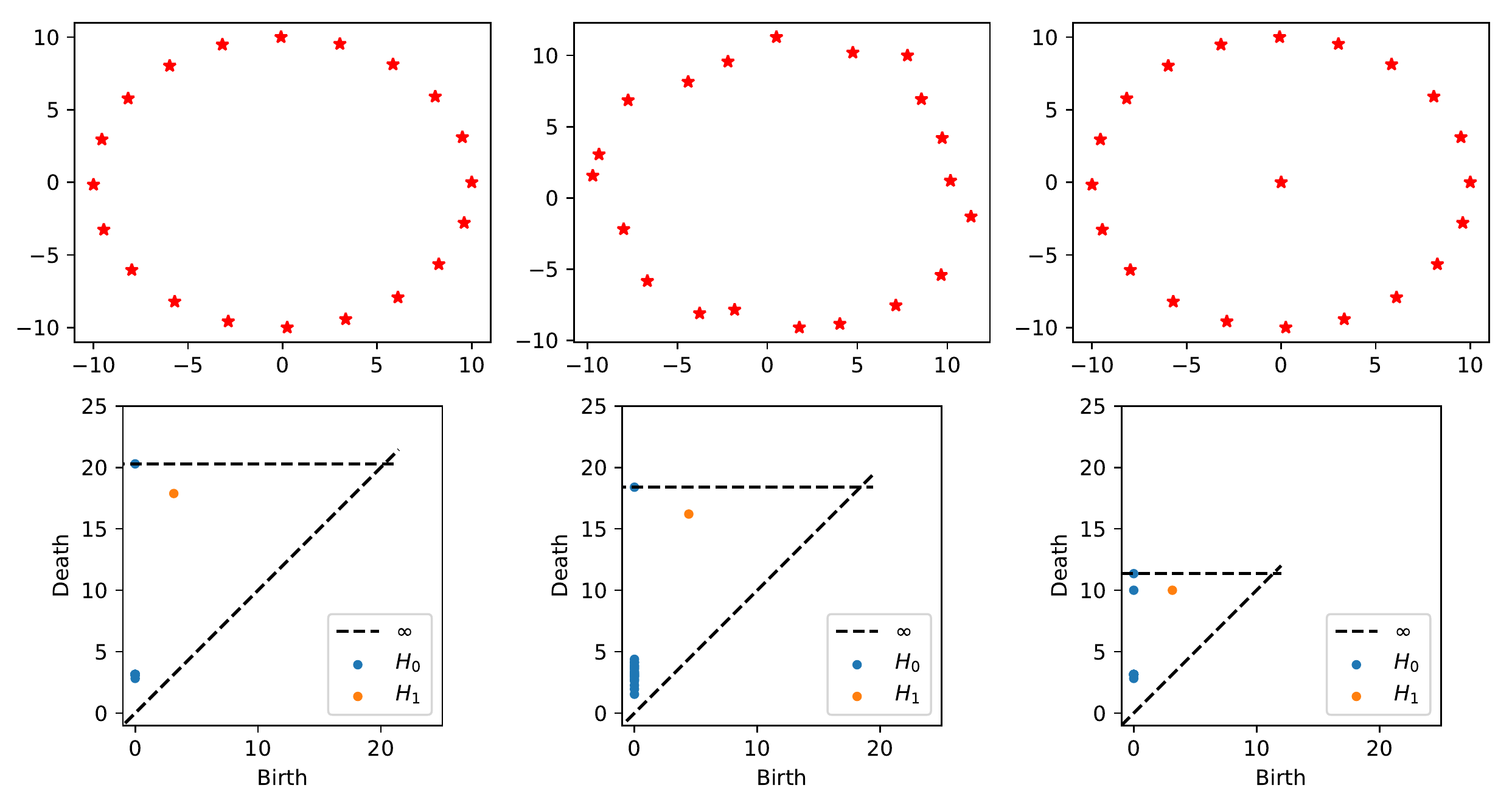}
    \caption{Persistence diagrams of clouds of points. The first row shows the cloud of point data: Original data sampled from a circle (LEFT), original data plus noise (MIDDLE) and finally, the original data plus an outlier in the centre (RIGHT). The second row displays the corresponding persistence diagram for each cloud of point. This figure demonstrates how much the persistence diagram can be sensitive to the presence of outliers. Thus, pointing out to the fact that we might need to transform the data before we can apply TDA.}
    \label{fig:stability_PD}
\end{figure}

A persistence diagram of a topological space is a multiset of points (birth-death pairs) that represent the various features present in the topology of the data set. A birth time means that at this specific time or scale a new topological feature appeared in the filtration, and the corresponding death time is the time or scale from which such topological feature is no longer present in the filtration. When dealing with one dimensional Morse functions we can only consider zero-dimensional features (number of connected components), but in general when analyzing an arbitrary set of points we can consider higher dimensional features (wholes, voids etc.) to better capture the shape of the underlying manifold at hand. For example, the persistence diagrams in Figure \ref{fig:stability_PD} show how even a little additive noise (with small $\sigma$) can affect the persistence diagram. Adding an outlier in the middle of the circle, however, changes completely the location of the dots (BLUE and ORANGE), which may lead to different conclusions about the the underlying unknown process that generated the data. Persistence diagrams help visualize/summarize the topological information contained in data. A detailed explanation of this notion will follow in the next section.

\subsection{Persistent homology of  Morse filtration}
\label{subsec:Morse_filtration}

Very often in time series analysis, the observed time series can be modeled using a mean structure plus a random (and possibly correlated) noise, as seen in Equation \ref{eq:additive_TS_model} below
\begin{align}
    y(t) = \mu(t) + \epsilon(t) \label{eq:additive_TS_model}
\end{align}
where the mean structure $ \mu(t) $ is supposed to capture the deterministic trend, and the noise $\epsilon(t)$ accounts for the stochastic fluctuations around the mean which captures the autocovariance (or generally, the within dependence) structure. Viewing $\mu(t)$ as a Morse function allows us to use Morse theory to build a sublevel set filtration that captures the topological information contained in $\mu$, specifically the arrangement of its critical values, see Theorem 3.20 in \cite{LECTURES_MORSE_HOMOLOGY} that says: The homotopy of the sublevel set only changes when the parameter value passes through a critical point. In a nutshell, the "homotopy" of a set considers only the critical information about its topology and disregards the effect of continuous deformations, for example, shrinking or twisting the set without tearing.

In general, EEG signals represent the superposition of numerous ongoing brain processes. Therefore, to be able to accurately characterize and estimate brain functional response to a stimulus, it is necessary to record many trials of the same repeated stimulus. In this case, a smoothing approach (i.e., average many time-locked single-trial EEGs recorded from the same stimulus) is meaningful as it allows random (nonrelated to the stimulus) brain activity to be canceled out and relevant signal to be enhanced. This new signal is referred to as event-related potential (ERP).

\begin{figure}
    \centering
    \includegraphics[width=\linewidth]{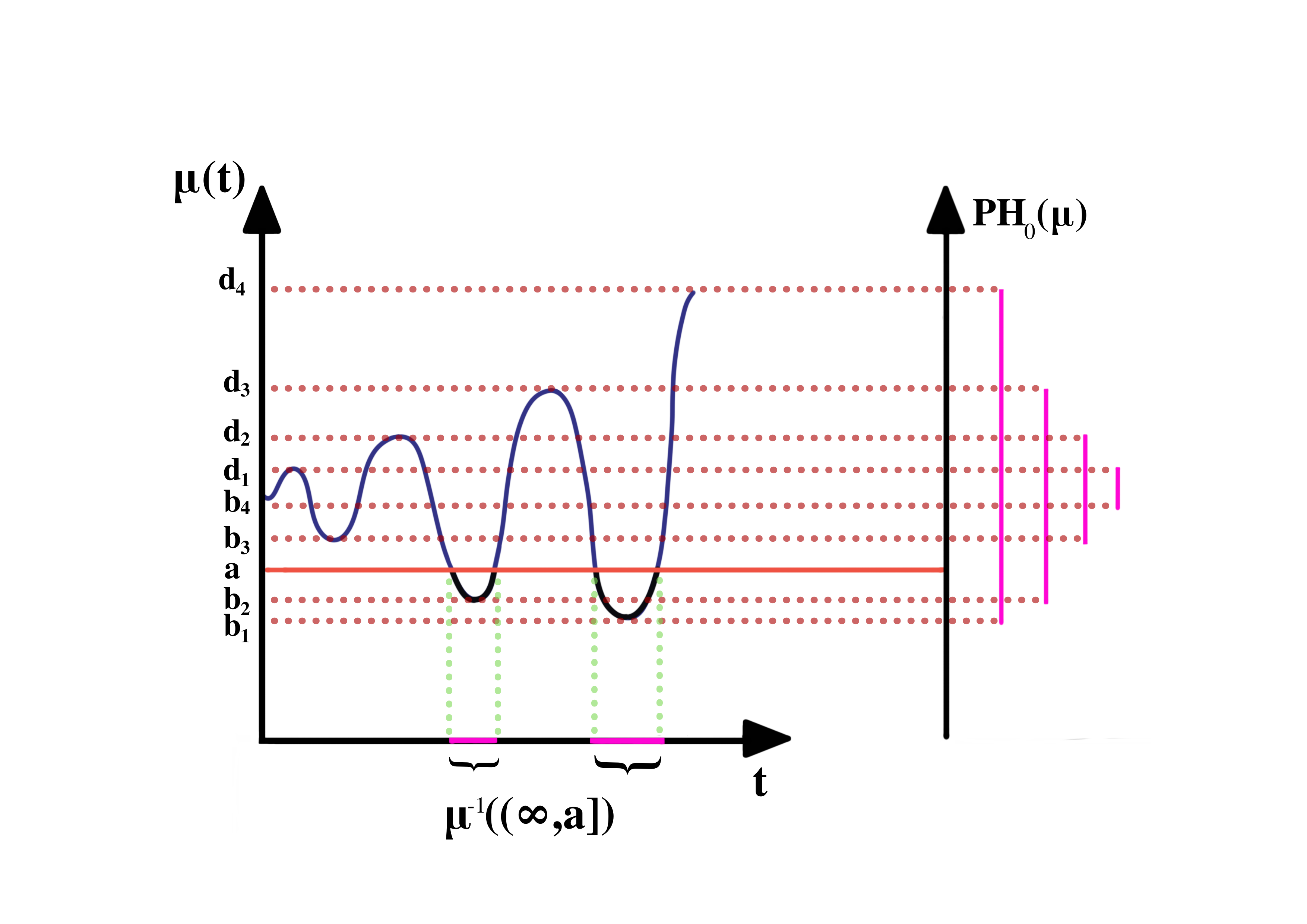}
    \caption{Sublevel set technique for a one-dimensional Morse function. On the left is the function $\mu(t)$, on the right is the barcode summary of the zero-dimensional persistence homology.}
    \label{fig:morse_filtration}
\end{figure}

Indeed, applying the Morse filtration to ERP data is meaningful as it allows to capture meaningful information regarding the critical values of the ERP signal. On account of the additive nature of the noise $\epsilon$, a first step is needed to smooth the time series (i.e., recover the mean structure $\mu(t)$). After smoothing, the Morse filtration can be built then visualized as in Figure \ref{fig:morse_filtration}.
\noindent As the parameter $a$ increases, the corresponding sequence of preimages (i.e., $\mu^{-1}(]-\infty, a])$) forms a sublevel set filtration. The topology of the preimage only changes when $a$ goes through a critical value with non vanishing second derivative or non singular Hessian matrix. At every critical value, a component is either born (at local minimums) or dies (at local maximums) by merging with another component.

\begin{figure}
    \centering
    \includegraphics[width=\linewidth]{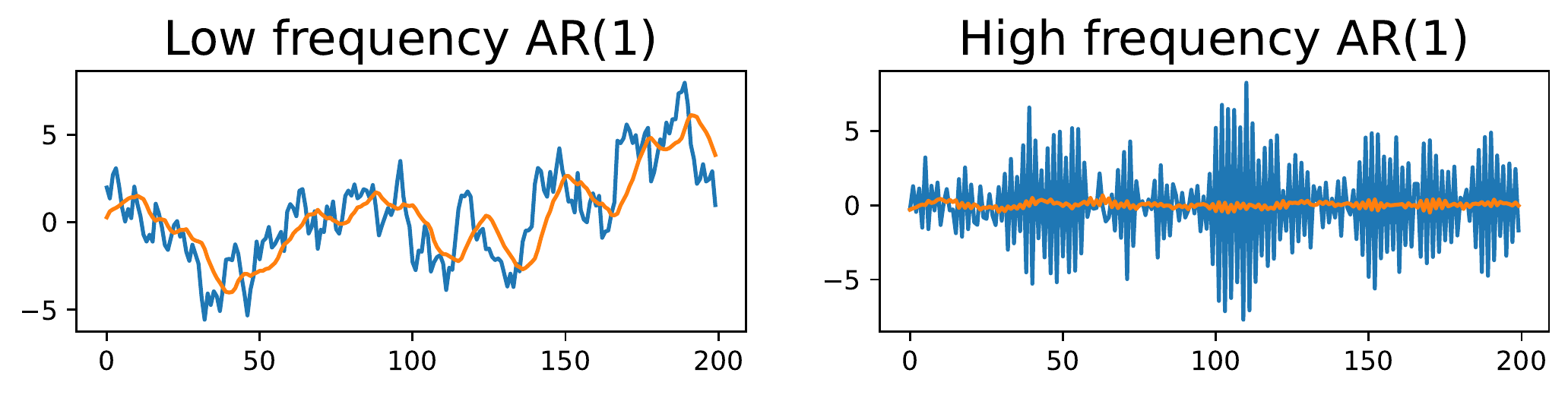}
    \caption{The left time series represents a zero mean plus a low-frequency AR(1) process ($\rho=0.95$), original time series in blue and smoothed time series in orange. The right time series represents a zero mean plus a high-frequency AR(1) process ($\rho=-0.95$), original time series in blue and smoothed time series in orange.}
    \label{fig:effect_of_smoothing}
\end{figure}

The Morse filtration summarizes the topological information contained in the mean of the time series, $\mu(t)$, by capturing the information contained in the arrangement of the local extrema. This implies that TDA ignores the information contained in the noise structure, e.g., the covariance and dependence structure. Therefore, the practitioner has to verify that the assumptions of Equation \ref{eq:additive_TS_model} are valid. Otherwise applying the Morse filtration will result in big information loss. For example, in time series analysis, such a model could have disastrous consequences, as in the case of autoregressive processes, see Figure \ref{fig:effect_of_smoothing}. In this example, the presence of a high-frequency AR(1) process does not modify the mean structure significantly, which does not alter the conclusions of the Morse filtration as the smoothing step cancels the noise structure and unravels the true mean structure. The presence of a low-frequency AR(1) noise process can be problematic because this can be incorrectly absorbed into the mean structure $\mu(t)$. This could lead the Morse filtration into erroneous conclusions.

\subsection{Persistent homology of Vietoris-Rips filtration}
\label{subsec:Persistent_homology}

\begin{figure}
    \centering
    \includegraphics[width=.8\linewidth]{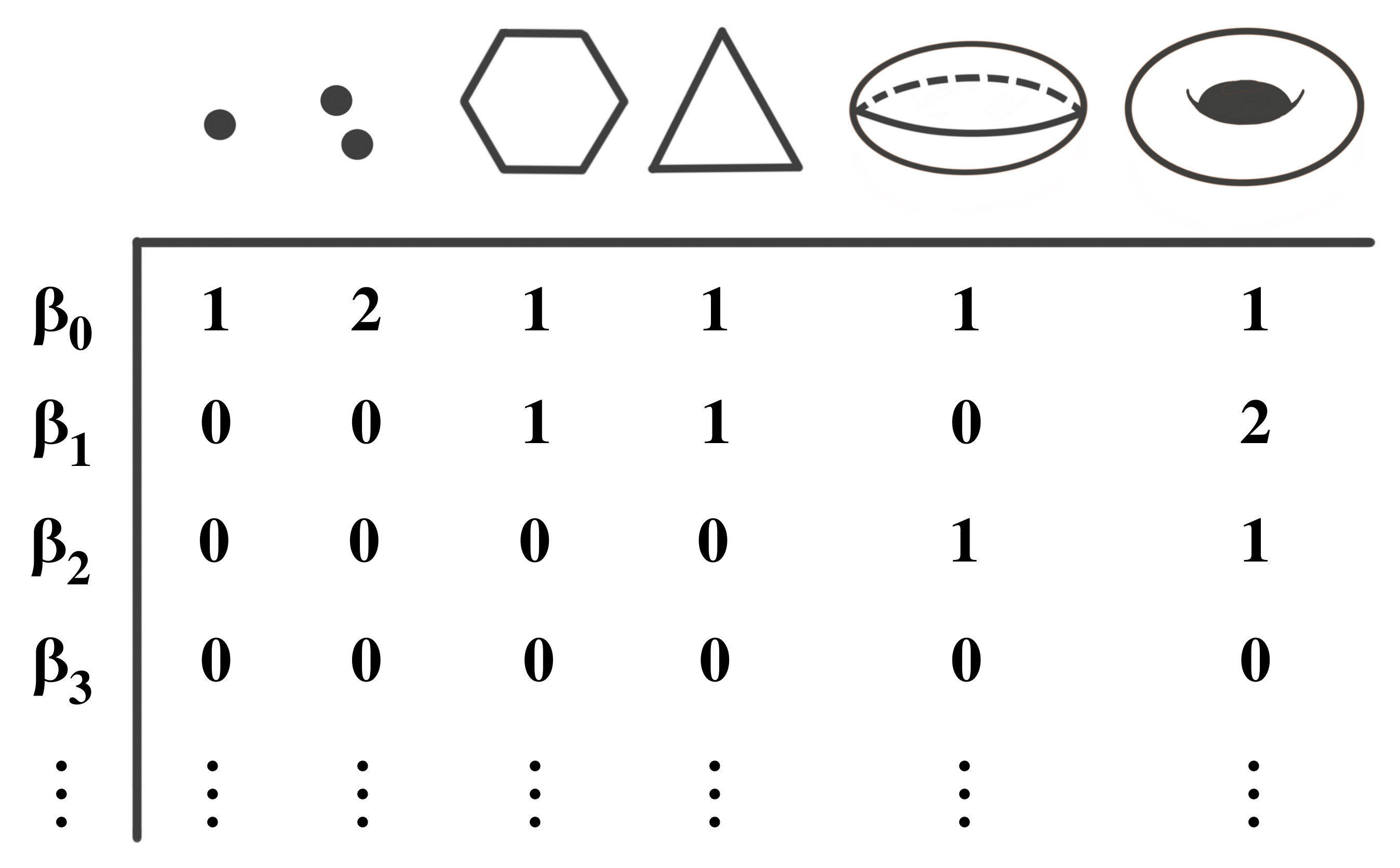}
    \caption{Examples of topological object with the corresponding Betti numbers. The zero-Betti number ($\beta_0$) counts the number of components, the one-Betti number ($\beta_1$) counts the number of cycles or wholes, the two-Betti number ($\beta_2$) counts the number of voids etc.}
    \label{fig:betti_number_illustration}
\end{figure}

\begin{figure}
    \centering
    \includegraphics[width=\linewidth]{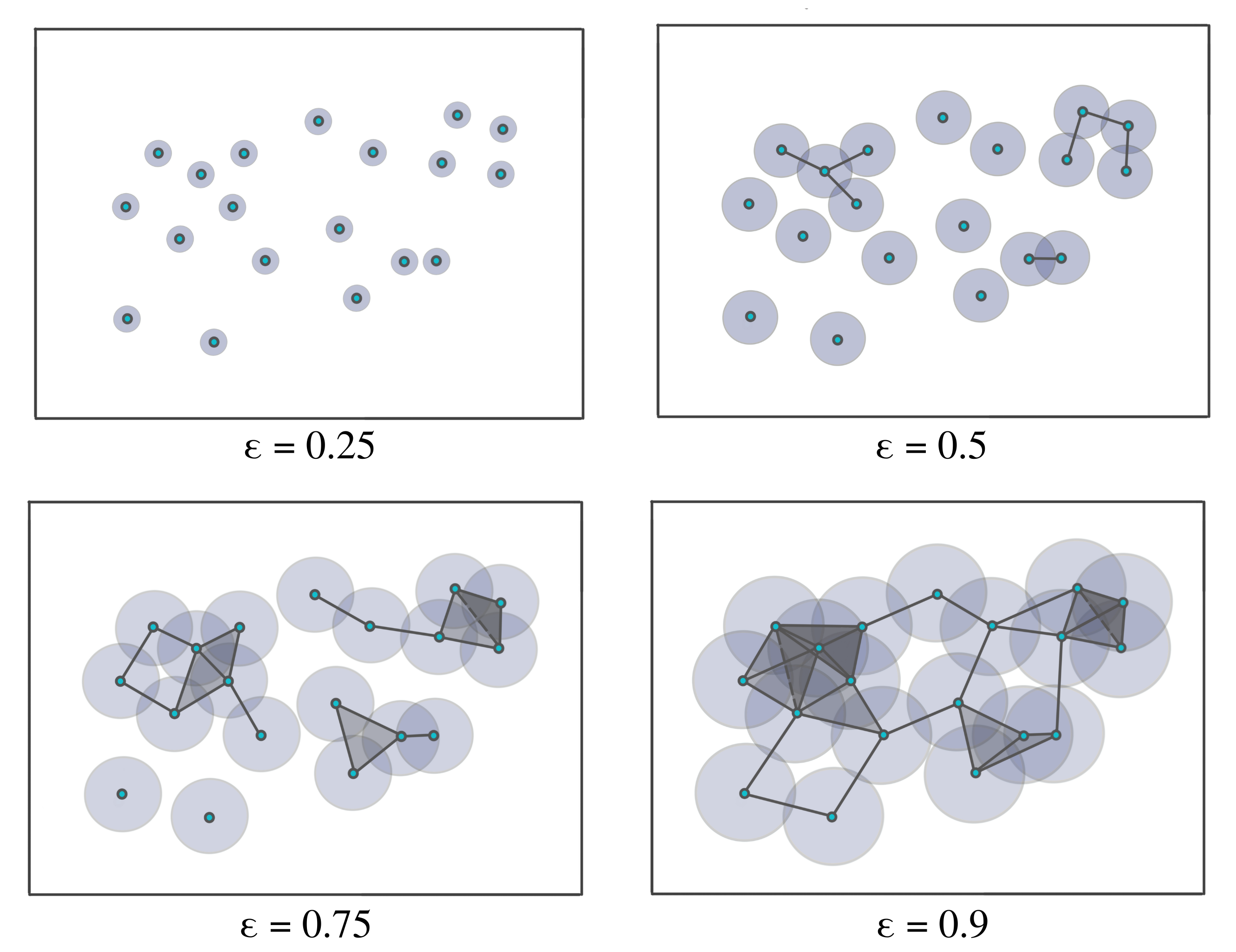}
    \caption{Example of a Vietoris-Rips filtration. As the radius $\epsilon$ increases, the balls centered around the data points with radius $\epsilon$ start intersecting, which leads to the appearance of more and more complex features in the increasing sequence of simplicial complexes.}
    \label{fig:PH_example}
\end{figure}

\begin{figure}
    \centering
    \includegraphics[width=\linewidth]{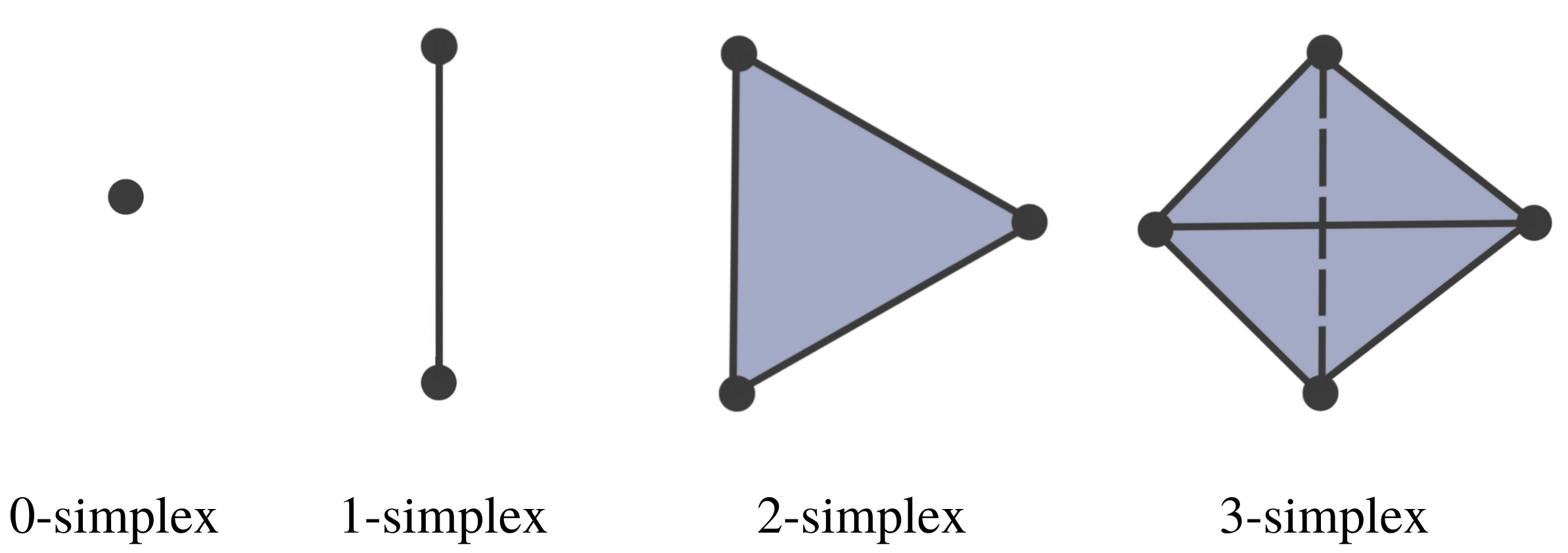}
    \caption{Simplices with different dimensions.}
    \label{fig:simplicial_complexes_example}
\end{figure}

\begin{figure}
    \centering
    \includegraphics[width=\linewidth]{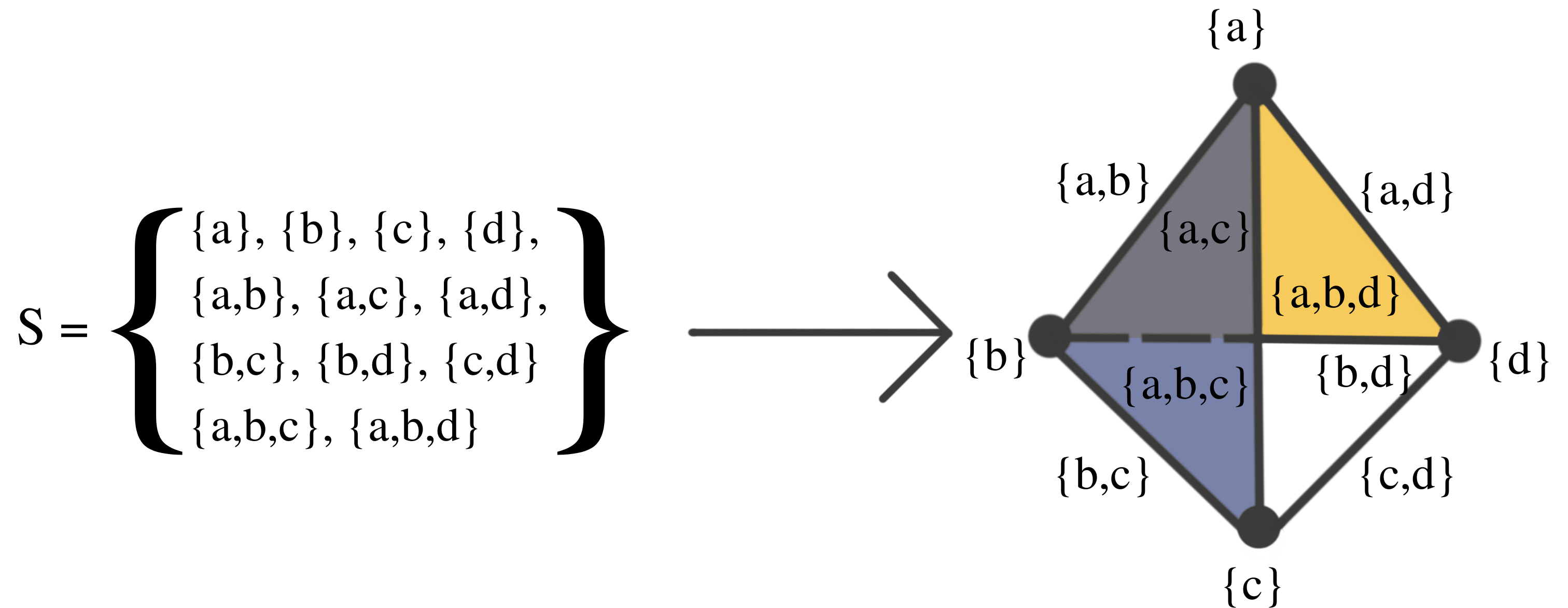}
    \caption{Example of a simplicial complex with four nodes, six edges and two faces. If $S$ is a simplicial complex, then every face of a simplex in $S$ must also be in $S$.}
    \label{fig:simplicial_complex_example}
\end{figure}

The original motivation behind homology theory is to distinguish between topological objects algebraically, using group theory based on Betti numbers (Figure \ref{fig:betti_number_illustration}). Homology is a tool from algebraic topology that analyzes the topological features of objects, such as connected components, holes, cavities, etc. For example, Figure \ref{fig:betti_number_illustration} presents a few topological objects (e.g., a sphere, a torus etc.) with various topological features of different dimensions such as wholes and cavities. Often, the data is assumed to be a finite set that is sampled with additive noise from an underlying topological space.
In order to analyze the shape of the data, we usually build the homology of the data by looking at the generated networks of neighboring data points at varying scales/distances, as seen in Figure \ref{fig:PH_example}. We call this sequence of increasing networks Vietoris-Rips filtration, see \cite{HAUSMANN_RIPS_FILTRATION}, as we increase the radius $\epsilon$ the balls surrounding the data points start intersecting and thus form a network of neighboring points, see Figure \ref{fig:PH_example}. The goal of such approach is to detect the nature of the geometrical patterns when they appear (birth) and for how long they persist (death time minus birth time) for a wide range of radius values.
The Vietoris-Rips filtration is constructed based on the notions of a simplex and simplicial complex, which is a finite collection of sets that is closed under the subset relation, see Figures \ref{fig:simplicial_complexes_example} and \ref{fig:simplicial_complex_example}. Simplicial complexes can be thought of as higher dimensional generalization of graphs. The use of simplicial complexes aims for summarizing the shape of the data 
at every scale by a mathematical object that simplifies abstract manipulations.
Simplicial complexes can be as simple as a combination of singleton sets (disconnected nodes), or more complicated such as a combination of pairs of connected nodes (edges), triplets of triangles (faces), quadruplets of tetrahedrons, or any higher dimensional simplex as in Figure \ref{fig:simplicial_complexes_example} or a combination of different k-simplices in general as in Figure \ref{fig:simplicial_complex_example}. The notion of a simplicial complex may seem almost identical to the notion of networks. However, there is a major difference. In spite of this, there is still a significant difference. On the one hand, networks and graphs disregard surfaces, volumes, etc., and can be thought of as flexible structures, whereas simplicial complexes take a much richer approach by keeping track of many levels of complexity represented by various k-simpleces. 

The above definition of simplicial complexes gives a rigorous description of the Vietoris-Rips filtration as an increasing sequence of simplicial complexes. To construct this increasing sequence of simplicial complexes, practitioners use the concept of a cover with an open ball around each node. An important motivation behind this approach is the Nerve theorem. Historically proposed by Pavel Alexandrov (sometimes attributed to Karol Borsuk), the Nerve theorem simplifies continuous topological spaces into abstract combinatorial structures (simplicial complexes) that preserve the underlying topological structure and can be examined by algorithms. Indeed, as described in \cite{NERVE_THEOREM}, the Nerve theorem states that a set and the nerve of the set covering are homotopy equivalent as the resolution of the cover increases, i.e., they have identical topological properties, such as the number of connected components, holes, cavities etc.
\begin{figure}
    \centering
    \includegraphics[width=\linewidth]{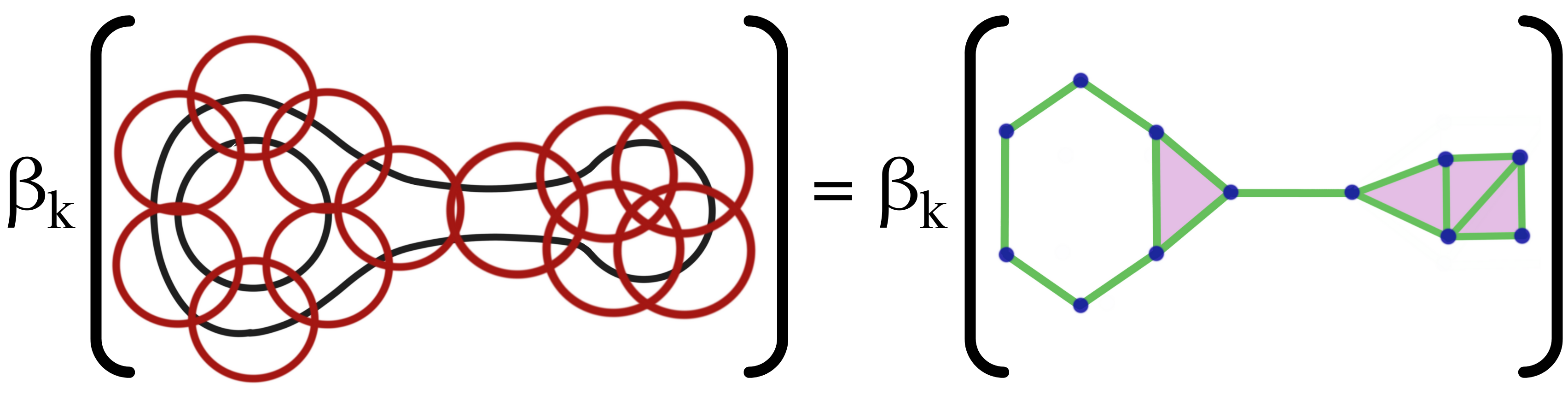}
    \caption{The open cover (LEFT) and the corresponding nerve (RIGHT) have identical Betti numbers, denoted by $\beta_k$ (i.e., number of connected components, holes, voids etc.). As the resolution of the cover increases the topological structure of the resulting nerve resembles that of the original space.}
    \label{fig:nerve_of_covering}
\end{figure}
As a result of the Nerve's theorem, see Figure \ref{fig:nerve_of_covering}, the topological properties (i.e., Betti numbers) of the simplicial complexes generated from the open cover should emulate those of the underlying manifold as the resolution of the covering increases, and thus the open cover converges to the original manifold.

\begin{figure}
    \centering
    \includegraphics[width=0.6\linewidth]{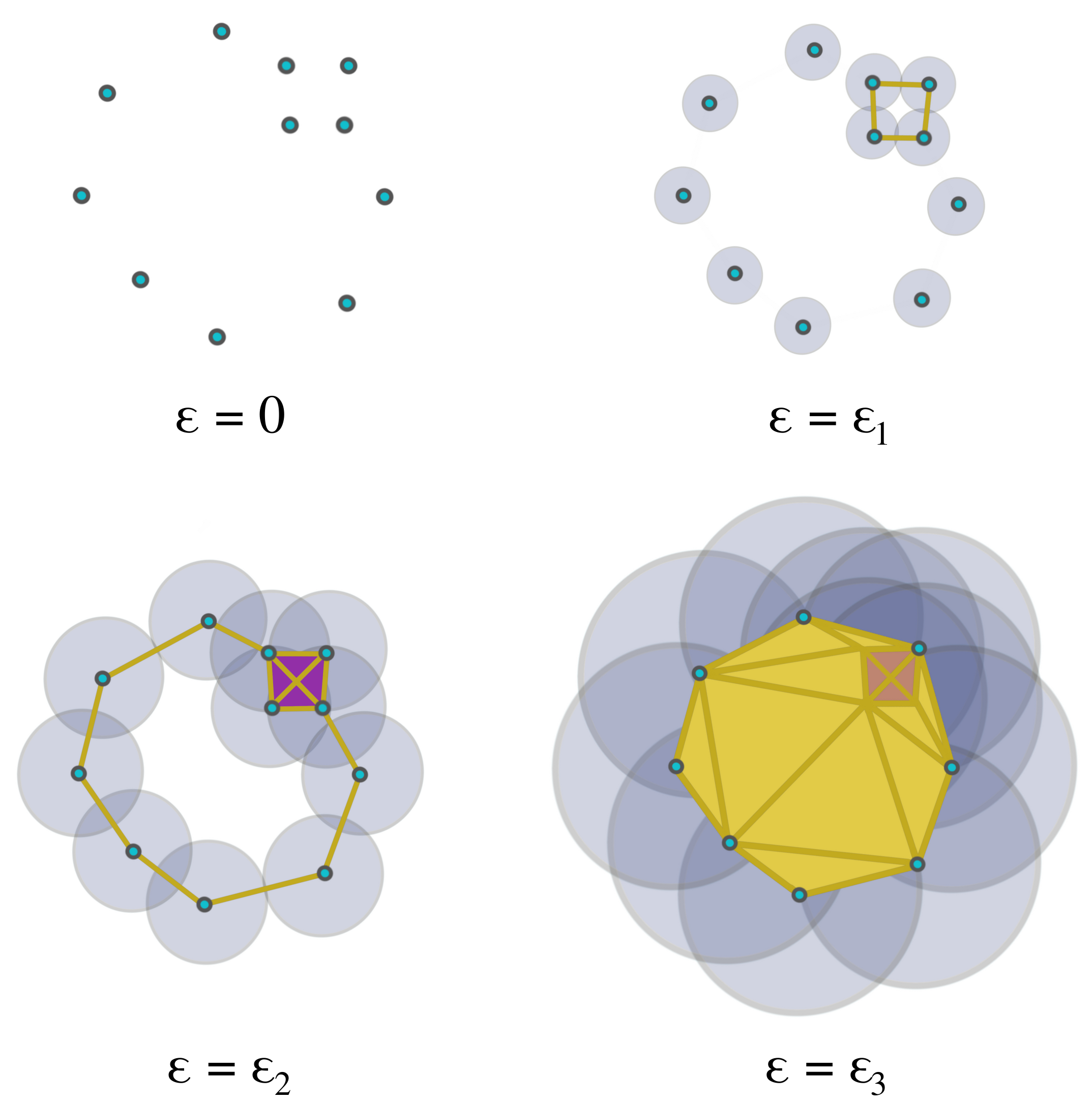}
    \caption{When $\epsilon=0$ all points are disconnected, as $\epsilon$ grows the open cover becomes larger and larger. When $\epsilon=\epsilon_1$ some of the balls start to intersect and thus an edge (or higher dimensional simplex, if more than two balls intersect) is created between such pair of points which results in the creation of a cycle (one dimensional whole). When $\epsilon=\epsilon_2$, more edges (ten 1-simpleces) get added and as well as a tetrahedron (four 2-simpleces and one 3-simplex) which results in the creation of a new cycle and destruction of the first cycle. Finally, when $\epsilon=\epsilon_3$ more simpleces are added and the second cycle disappears. This Figure was inspired form the paper by \cite{USER_GUIDE_TDA}.}
    \label{fig:filtration_example}
\end{figure}

\begin{figure}
    \centering
    \includegraphics[width=\linewidth]{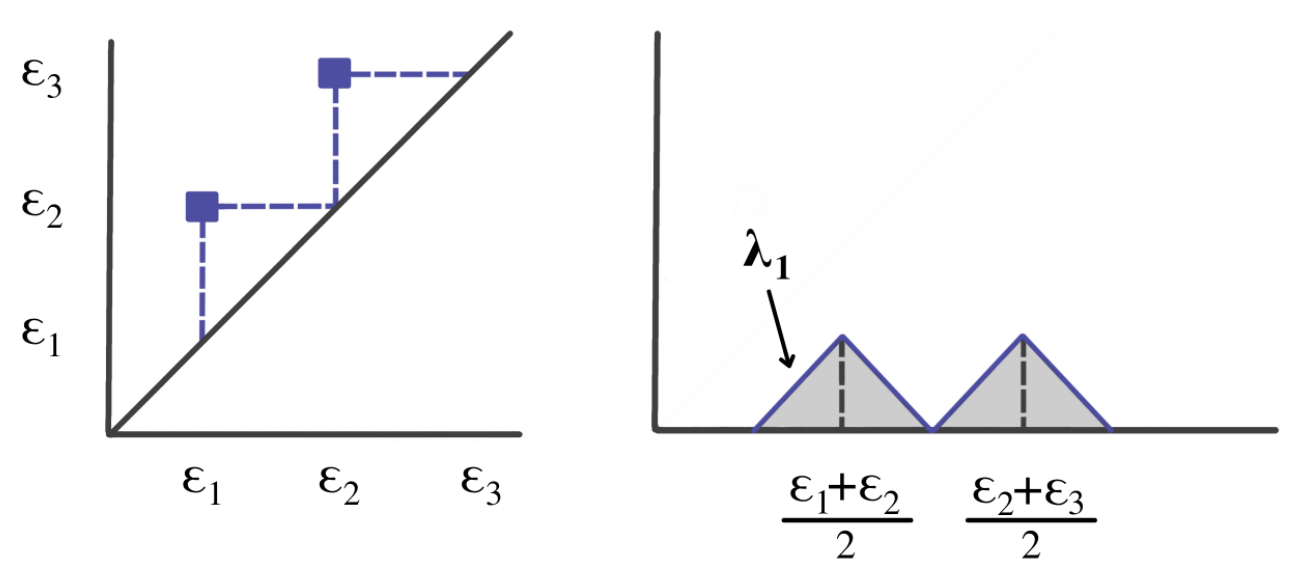}
    \caption{Corresponding persistence diagram (LEFT) and persistence landscape (RIGHT) from the cloud of points defined in Figure \ref{fig:filtration_example}. This Figure is inspired form the paper by \cite{TDA_GUIDEA}.}
    \label{fig:PD_PL_example_1}
\end{figure}

\begin{figure}[b]
    \centering
    \includegraphics[width=\linewidth]{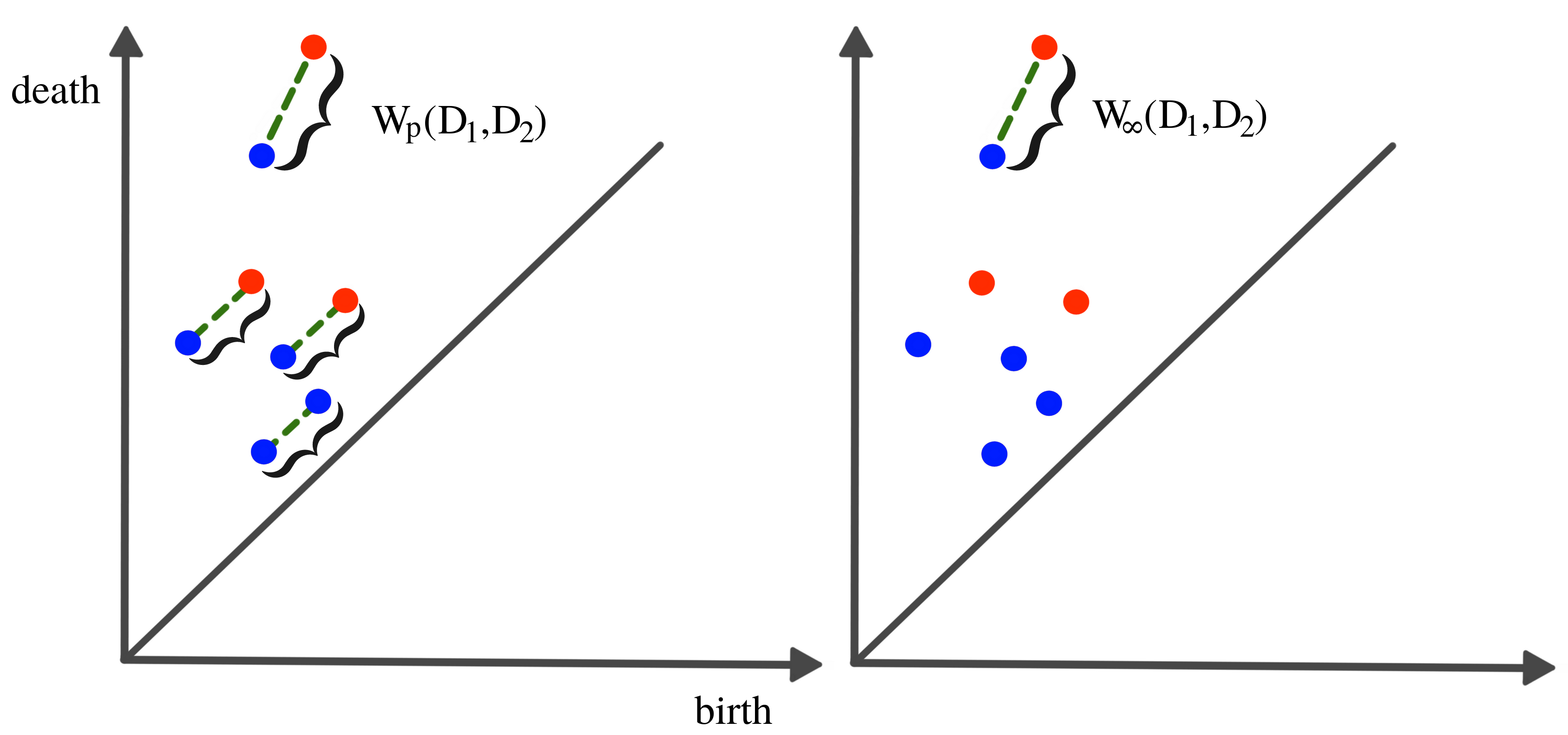}
    \caption{Wasserstein distance (LEFT) vs Bottleneck distance (RIGHT). To compute any of the two distances the optimal point correspondence needs to be found which might become computationally infeasible as the number mappings increases in the order of $\mathcal{O}(n_E^{n_F})$.}
    \label{fig:bottleneck_wassetstein_distance}
\end{figure}

\begin{figure}[b]
    \centering
    \includegraphics[width=\linewidth]{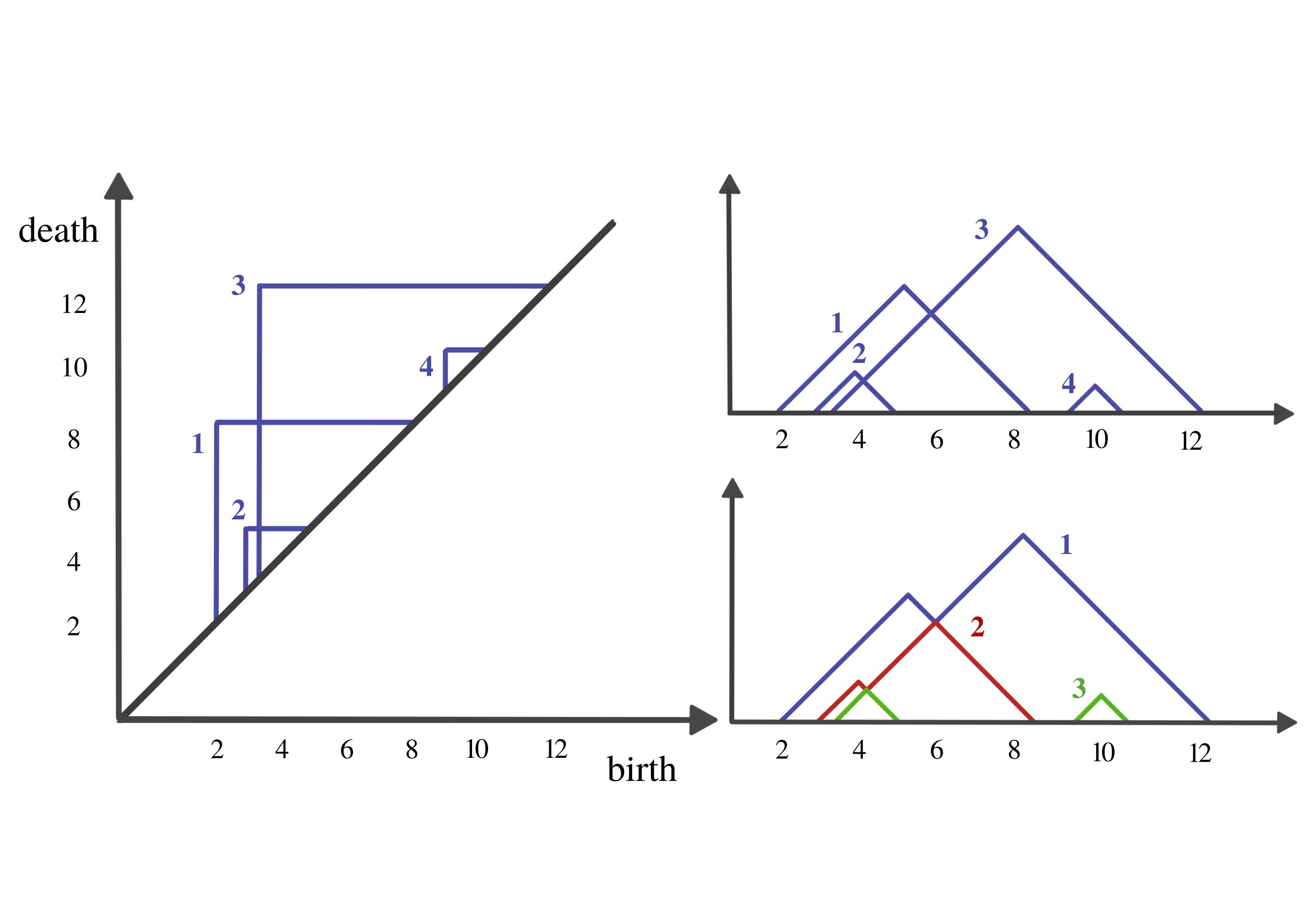}
    \caption{Construction of a complex persistence landscape (PL) from a persistence diagram (PD). This Figure is inspired form the paper by \cite{PL_FIRST}.}
    \label{fig:PD_PL_example_2}
\end{figure}

In practice, to construct the Vietoris-Rips filtration from a finite set of points, one looks at the increasing finite cover ($\bigcup_{i=1}^n \mathcal{U}_i(r)$, where $\mathcal{U}_i(r)$ is a ball centered around the $i^{th}$ point/node with radius $r$) of the topological space of interest at a wide range of radius values, as seen in \ref{fig:simplicial_complexes_example}. Another example is considered in Figure \ref{fig:filtration_example} where the threshold values where topological features appear or disappear are denoted by $\epsilon_i$. Once the persistent homology is constructed, it needs to be analyzed using some topological summary such as the barcode, the persistence diagram or the persistence landscape. In Figure \ref{fig:PD_PL_example_1} we see the representation of the corresponding persistence diagram and the persistence landscape.
The persistence diagram (PD) is constructed based on the times of birth and death of the topological features in the filtration as seen in Figure \ref{fig:PD_PL_example_1}. Thus, for every birth-death pair a point is represented in the diagram, e.g., ($\epsilon_1$, $\epsilon_2$) and ($\epsilon_2$, $\epsilon_3$). The points in the PD are color coded so that every color represents a specific dimension of the homology (dimension 0 for connected components, dimension 1 for cycles etc.).
It is hard to manipulate (take averages or compute distances) persistence diagrams (e.g., compute Bottleneck distance or Wasserstein distance, see Figure \ref{fig:bottleneck_wassetstein_distance}). In \cite{WASSERSTEIN_BOTELLNECK}, the authors compare persistence diagrams. In particular, they show how it can be time consuming to compute the Bottleneck or the Wasserstein distances as it is necessary find point correspondence. Additionally, defining a "mean" (or center) or the "variation" (or measure of spread) of a distribution  persistence diagrams is not straightforward, especially when the number of points in each diagram varies.
For these reasons, practitioners prefer to analyze a transformation (persistence landscape) of the persistence diagram, which is a simpler object (function), as defined in \cite{PL_FIRST}, see Figure \ref{fig:PD_PL_example_1}. The persistence landscape (PL) can be constructed from the persistence diagram (PD) by drawing isosceles triangle for every point of a given homology dimension in the PD centered around the birth and death times, as shown in Figure \ref{fig:PD_PL_example_1}. In case there are intersecting lines, the most persistent (highest) function is defined to be the PL (i.e., $\lambda_1$), see example in Figure \ref{fig:PD_PL_example_2}. For more details regarding the properties of the persistence landscape refer to \cite{PL_FIRST}.
Since the PLs are functions of a real variable (scale), it becomes very easy to compute group averages and to define confidence regions, and all sorts of statistical properties (such as the strong law of large numbers, or the central limit theorem) could be derived for the PL, as shown in \cite{PL_FIRST}.

\subsection{Time delay embeddings of univariate time series}
\label{subsec:Embedding_technique}

\begin{figure}[b]
    \centering
    \includegraphics[width=\linewidth]{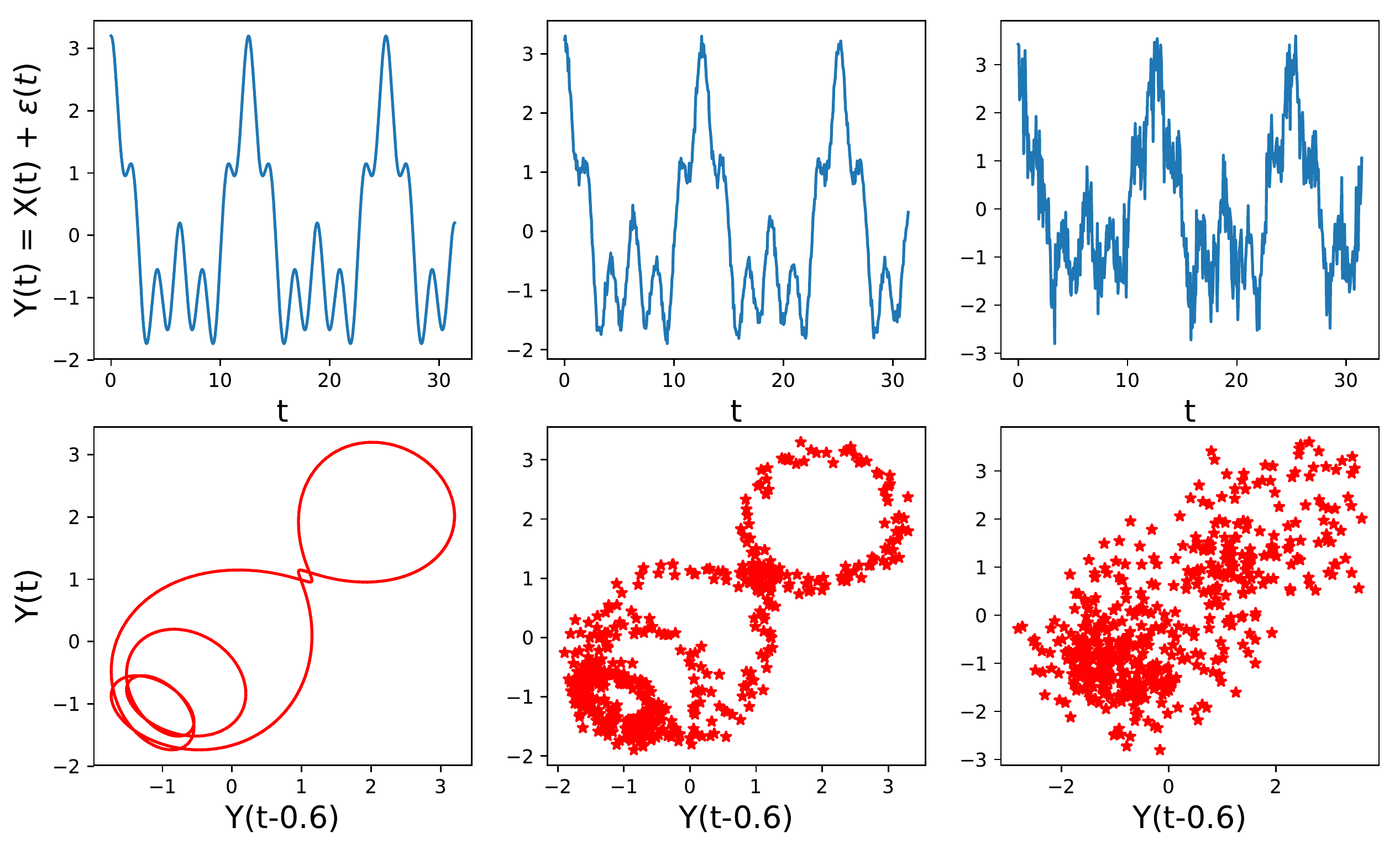}
    \caption{Point cloud embedding of a univariate time series using sliding window method for different noise levels. Original time series (TOP), for various standard deviations of the noise $\epsilon$ (from left to right $\sigma=0$, $\sigma=0.1$, $\sigma=0.5$), with corresponding time delay embedding (BOTTOM). The dependence structure in this time delay embedding cannot be visually observed even in the presence of moderate level of noise.}
    \label{fig:time_series_embedding_illustration_2}
\end{figure}

So far we described the process of building the persistence homology from a cloud of points living in a metric space. However, in order to analyze time series data using the previous method, it is necessary to create some kind of embedding of the univariate time series into a metric space. For instance, instead of studying the time series $\{Y(t), t=1, \ldots, T\}$, we will study the behaviour of the cloud of points $\{ Z_s = (Y(s), Y(s-1)), s=2, \ldots, T\}$, as seen in Figure \ref{fig:time_series_embedding_illustration_2}. This particular embedding is known as the time delay embedding, see \cite{TIME_DELAY_TAKENS}, and it aims at reconstructing the dynamics of the time series by taking into consideration the information in lagged observations. 

\begin{figure}[b]
    \centering
    \includegraphics[width=.8\linewidth]{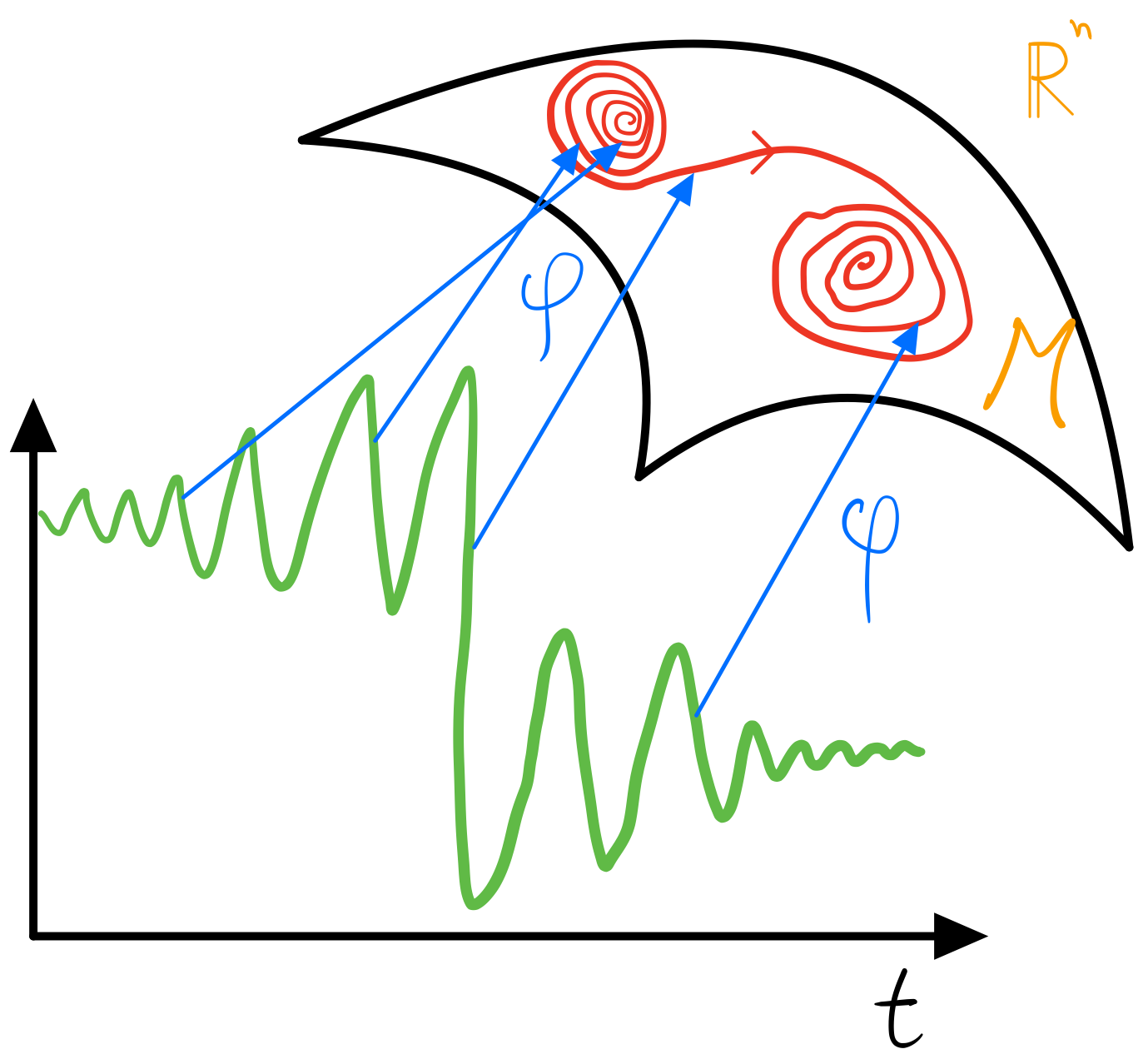}
    \caption{Illustration of the reconstructed (using the embedding map $\phi$) attractor (red curve in the manifold $\mathcal{M}\subset \mathbf{R}^n$) from the time series observations.}
    \label{fig:time_series_embedding_illustration_1}
\end{figure}

In a time delay embedding, the aim is to reconstruct the phase space (the space that represents all possible states of the system) based on only one observed time series component, borrowing information from the lagged observations to to do so. Under the initial assumptions, this is indeed possible since all the components are interdependent through the shared dynamical system. This phase space may contain valuable information regarding the behaviour of the time series. For example, the time series might display some chaotic behaviour in time, see \ref{fig:time_series_embedding_illustration_1}, however, in the phase space it might show some convergence to a strange attractor (a region of the phase space towards which the system converges), see \cite{TIME_DELAY_TAKENS} and \cite{TS_TDA} for more details regarding the application of topological data analysis to time series data. The foundation of this method derives from the framework of dynamical systems. Indeed, Takens's theorem states conditions under which the attractor of a dynamical system can be reconstructed from a sequence of observations, as it can be seen in Figure \ref{fig:time_series_embedding_illustration_1}.

Therefore, when practitioners use topological data analysis to analyse the shape of the point cloud embedding of a time series, they are in fact assessing the geometry of the attractor of the underlying dynamical system.
This approach is perfectly meaningful if the initial assumptions of the Takens's theorem are valid, that is assuming there is a corresponding dynamical system that is continuous and invertible that corresponds to the observed time series. However, very often in time series analysis such approach does not make sense, especially when the observed time series is noisy, has constant mean and has no apparent relation with any dynamical system and hence it is unlikely to observe such patterns. For example in Figure \ref{fig:time_series_embedding_illustration_2}, we see that the time embedding does not show any interesting geometrical features when the level of the noise is high.

\section{Topological methods for analyzing multivariate time series}
\label{sec:TDA_MV_TS}

Over the last three decades, the analysis of the human connectome using various brain imaging techniques such as functional magnetic resonance imaging (fMRI) and electroencephalography (EEG) has witnessed numerous successes (see \cite{BOOK_FMRI_LAZAR}, \cite{FMRI_LINDQUIST}, \cite{FMRI_PAIN}, \cite{EEG_BRAIN_NETWORK}, \cite{EEG_EXPERIMENT}, \cite{RELIABLE_FUNCTIONAL_CON}, \cite{STAT_MODELS_EVO}, \cite{DYN_CON_fMRI}, \cite{DYNAMIC_COMUNITY_STRUCTURE}, \cite{EXTREMAL_CON}) to discover the background mechanisms of human cognition and neurological disorders. In this regard, the analysis of the dependence network of a multivariate time series from a topological point of view will have the potential to provide valuable insight.

\begin{figure}[b]
    \centering
    \includegraphics[width=\linewidth]{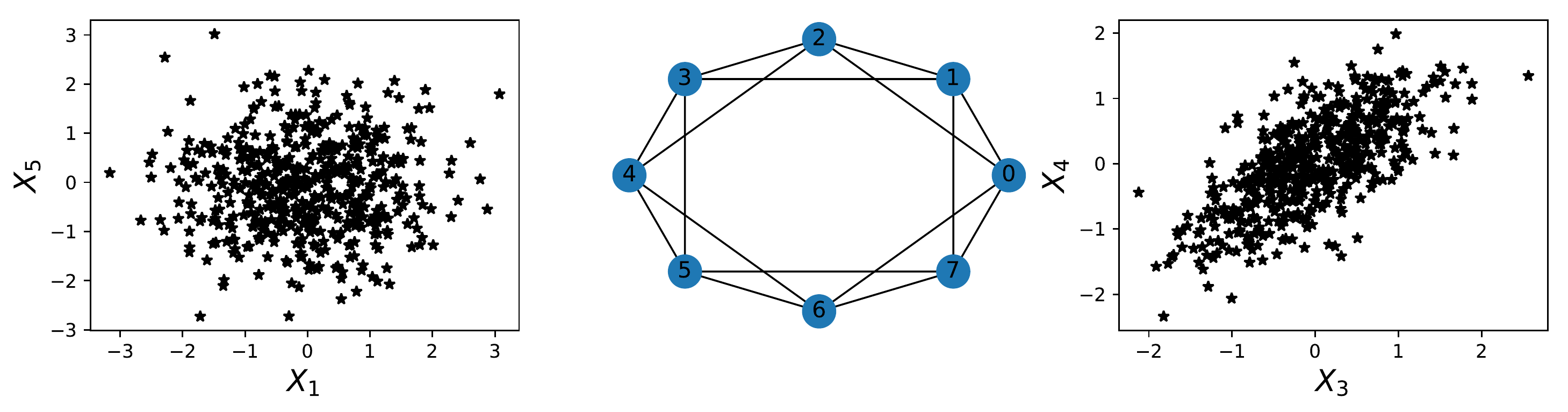}
    \caption{Illustration of a hidden cyclic pattern in the dependence structure. Left subplot, shows a scatter plot between time series components $X_5$ and $X_1$, right subplot, shows a scatter plot between time series components $X_4$ and $X_3$. Whereas, the subplot in the middle shows the cyclic latent dependence network of the entire multivariate time series.}
    \label{fig:hidden_topological_pattern_illustration}
\end{figure}

A multivariate time series might not display any relevant geometrical features in the point cloud embedding. However, it might display topological patterns in its dependence network as seen in Figure \ref{fig:hidden_topological_pattern_illustration}.
For this example, none of the previous methods can capture the interdependence between different time series components. However, an application of TDA to the time series' dependence network would directly reveal the cyclic pattern.

Traditionally, due to their stochastic nature, multichannel or multivariate brain signals  have often been modeled using their underlying dependence network, i.e., dependence between brain regions or nodes in a brain network, see \cite{BRAIN_ACTIVITY_NETWORK_REST}, \cite{MULTICHANNEL_BRAIN_SIGNALS}, \cite{SPECTRAL_DEPENDENCE}, \cite{EEG_BRAIN_NETWORK}. The brain network is often constructed starting from some connectivity measure derived from the observed brain signals. There are many possible characterizations of dependence and hence many possible connectivity matrices. These include cross-correlation, coherence, partial coherence, partial directed coherence, see \cite{SPECTRAL_DEPENDENCE}.

Due to the difficulty to analyze and visualize weighted networks, very often, practitioners make use of some thresholding techniques to create a binary network (from the weighted brain network) where the edge exists when the continuous connectivity value exceeds the threshold, see \cite{TDA_BRAIN}, \cite{PROBLEM_THRESHOLDING_SW_NETWORKS}, \cite{OPTIMAL_THRESHOLDING} and \cite{TDA_BRAIN_PROS_AND_CONS}. This thresholding is usually done at an arbitrary level. A major problem associated with this common approach is the lack of principled criteria for choosing the appropriate threshold. Moreover, the binary network derived from  a single threshold might not fully characterize the dependence structure. Obviously, the thresholding approach induces a bias and a huge loss of information that produces a simplified network.

Consequently, applying topological data analysis directly to the original data (i.e., consider a filtration of connectivity graphs) appears as an appealing alternative to the arbitrary thresholding of weighted networks. First, the TDA approach detects all potential topological patterns present in the connectivity network. Second, by considering all possible thresholding values, TDA avoids the arbitrary thresholding problem. Refer to \cite{TDA_BRAIN} for more details regarding this problem.

To formalize these ideas, let $X_i(t)$ be a time series of brain activity at location $i \in V$ and time $t t \in \{1, ...T \}$, where $V=\{1, \hdots, P\}$ is the set of all $P$ sensor locations (in EEGs) or brain regions (in fMRI). Therefore, considering a set of $P$ brain channels (e.g., electrodes/tetrodes) indexed by $V$, the object $\mathcal{X} = (V, \mathcal{D})$ is a metric space, where $\mathcal{D}_{ij}$ is the dependence-based distance between channel $i$ and channel $j$ (i.e., between $X_i(t)$ and $X_j(t)$). We build the Vietoris-Rips filtration by connecting nodes of $\mathcal{X}$ that have a distance less or equal to some given $\epsilon$, which results in the following filtration:
\begin{align}
    \mathcal{X}_{\epsilon_1} \subset \mathcal{X}_{\epsilon_2} \subset \cdots \subset \mathcal{X}_{\epsilon_n}, \label{eq:filtration}
\end{align}
where $\epsilon_1 < \epsilon_2 < \cdots < \epsilon_{n-1} < \epsilon_n$ are the distance thresholds. Nodes within some given distance $\epsilon_i$ are connected to form different simplicial complexes, $\mathcal{X}_{\epsilon_1}$ is the first simplicial complex (single nodes) and $\mathcal{X}_{\epsilon_n}$ is the last simplicial complex (all nodes connected, i.e., a clique of size $n$). In general, $\mathcal{X}\epsilon$ for a given $\epsilon$ represents the simplicial complex thresholded at distance $\epsilon$. However, $\mathcal{X}\epsilon$ only changes for a finite number of distance values, those present in the distance function, i.e., there are at most $n=P(P-1)/2$ simplicial complexes in the filtration (this is the number of simplicial complexes in the filtration, of course the number of possible simplicial complexes must be much higher: $2^{\frac{P(P-1)}{2}}$). For a detailed review of how to build the Vietoris-Rips filtration based on a metric space refer to \cite{HAUSMANN_RIPS_FILTRATION}.

Given a topological object $\mathcal{X}$ with some filtration as defined in Equation \ref{eq:filtration}, the corresponding homology analyzes the object $\mathcal{X}$ by examining its $k$-dimensional holes through the $k$-th homology groups $H_k(\mathcal{X})$. The zero-dimensional holes represent the connected components or the clustering information,
the one-dimensional holes the represent loops, the two-dimensional holes represent the voids etc. The rank $\beta_k$ of $H_k(\mathcal{X})$ is known as the $k$-th Betti number, see illustration in Figure \ref{fig:betti_number_illustration}. Refer to \cite{MUNKRES_ALG_TOP} and \cite{MERKULOV_ALG_TOP} for more rigorous definitions of these topological objects.

\subsection{Examples of time series models}

\begin{figure}[b]
    \centering
    \includegraphics[width=\linewidth]{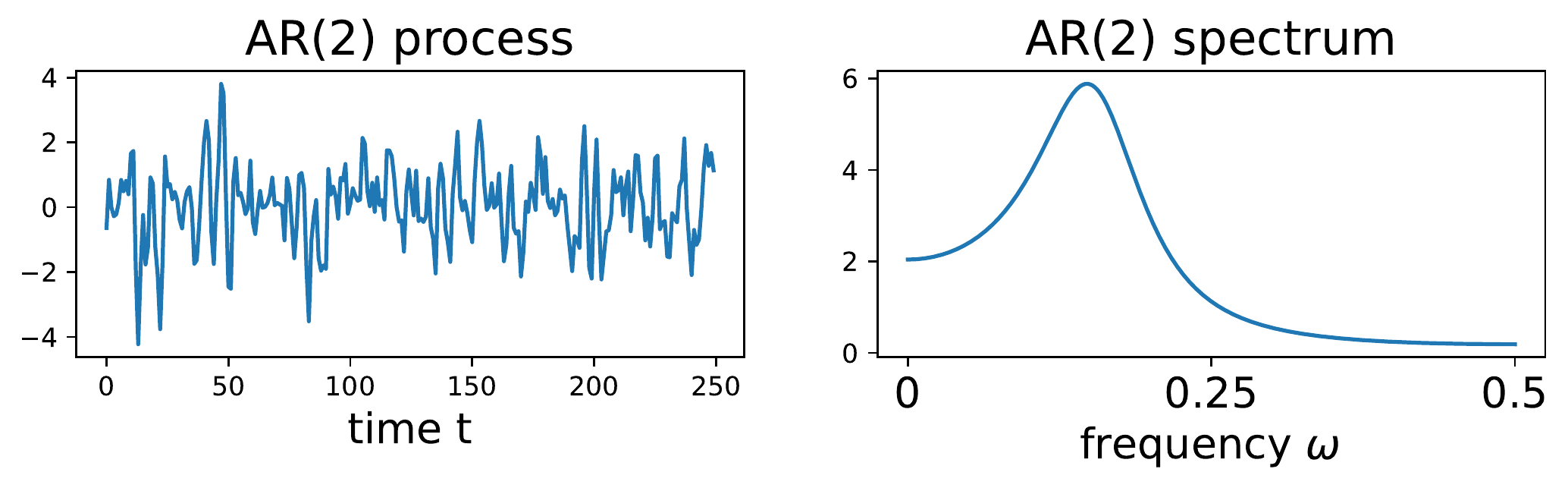}
    \caption{AR(2) process (LEFT) with $\phi_1 = \frac{2}{1.414} \cos(\pi \frac{154}{500})$ and $\phi_2 = - 1 / 1.414^2$ and its corresponding spectrum (RIGHT).}
    \label{fig:AR2_time_series_and_spectrum}
\end{figure}

In order to show how TDA can be used to analyze the dependence pattern of a real multivariate time series, we propose to illustrate via simulations how topological patterns such as cycles and holes can arise in multivariate time series. Based on the idea developed in \cite{SPECTRAL_DEPENDENCE} and \cite{AR2_MIXTURE} that recorded brain signals could be viewed as mixtures of latent frequency specific sources, i.e., mixture of frequency specific neural oscillations. These oscillatory activities can be modelled by a linear mixture of second order autoregressive (AR(2)) processes. A latent process with spectral peak at the alpha band (8-12Hz) can be characterized by an AR(2) process of the form:
\begin{align}
    Z^{\alpha}(t) &= \phi^{\alpha}_1 Z^{\alpha}(t-1) + \phi^{\alpha}_2 Z^{\alpha}(t-2) + W^{\alpha}(t) \label{Eq:Alpha_AR2}
\end{align}
where $W^{\alpha}(t)$ is white noise with $\mathbf{E} \hspace{1mm} W^{\alpha}(t) = 0$ and $Var \hspace{1mm} W^{\alpha}(t) = \sigma_\alpha^2$; and the AR(2) coefficients $\phi^{\alpha}_1$ and $\phi^{\alpha}_2$ are derived as follows. Note that Equation \ref{Eq:Alpha_AR2} can be rewritten as $W^\alpha(t) = (1 - \phi^{\alpha}_1 B^1 - \phi^{\alpha}_2 B^2)Z^{\alpha}(t)$ where the operator $B^k Z^{\alpha}(t) = Z^{\alpha}(t-k)$ for $k=1, 2$. The AR(2) characteristic polynomial function is:
\begin{align}
    \Phi(r) = 1 - \phi^{\alpha}_1 r^1 - \phi^{\alpha}_2 r^2.
\end{align}
Consider the case when the roots of the $\Phi(r)$, denoted by $r_1$ and $r_2$, are (non-real) complex-valued and hence can be expressed as $r_1 = M \exp(i 2 \pi \psi)$ and $r_2 = M \exp(-i 2 \pi \psi)$ where the phase $\psi \in (0, 0.5)$ and the magnitude $M>1$ to satisfy causality, see \cite{TSA_SHUMWAY_STOFFER}. For this latent process $Z^{\alpha}(t)$, suppose that the sampling rate is denoted by $SR$ and the peak frequency is $f_\alpha \in$(8-12Hz). Then the roots of the AR(2) latent process are $r^\alpha_1 = M_\alpha \exp(i 2 \pi \psi_\alpha)$ and $r^\alpha_2 = M_\alpha \exp(-i 2 \pi \psi_\alpha)$ where the phase $\psi_\alpha = f_\alpha / SR$. In practice, we can choose $\psi_\alpha = 10 / 100$ for a given $SR=100Hz$ and the root magnitude is $M_\alpha$ or some number greater than 1 but "close" to 1 so that the spectrum of $Z^{\alpha}(t)$ is mostly concentrated on the alpha band (8-12Hz). The corresponding AR(2) coefficients are derived to be $\phi_1^\alpha = \frac{2 \cos (2 \pi \psi_\alpha)}{M_\alpha}$ and $\phi_2^\alpha = -\frac{1}{M_\alpha^2}$. An example of such stationary AR(2) process can be visualized in Figure \ref{fig:AR2_time_series_and_spectrum}.

\begin{figure}[b]
    \centering
    \includegraphics[width=.7\linewidth]{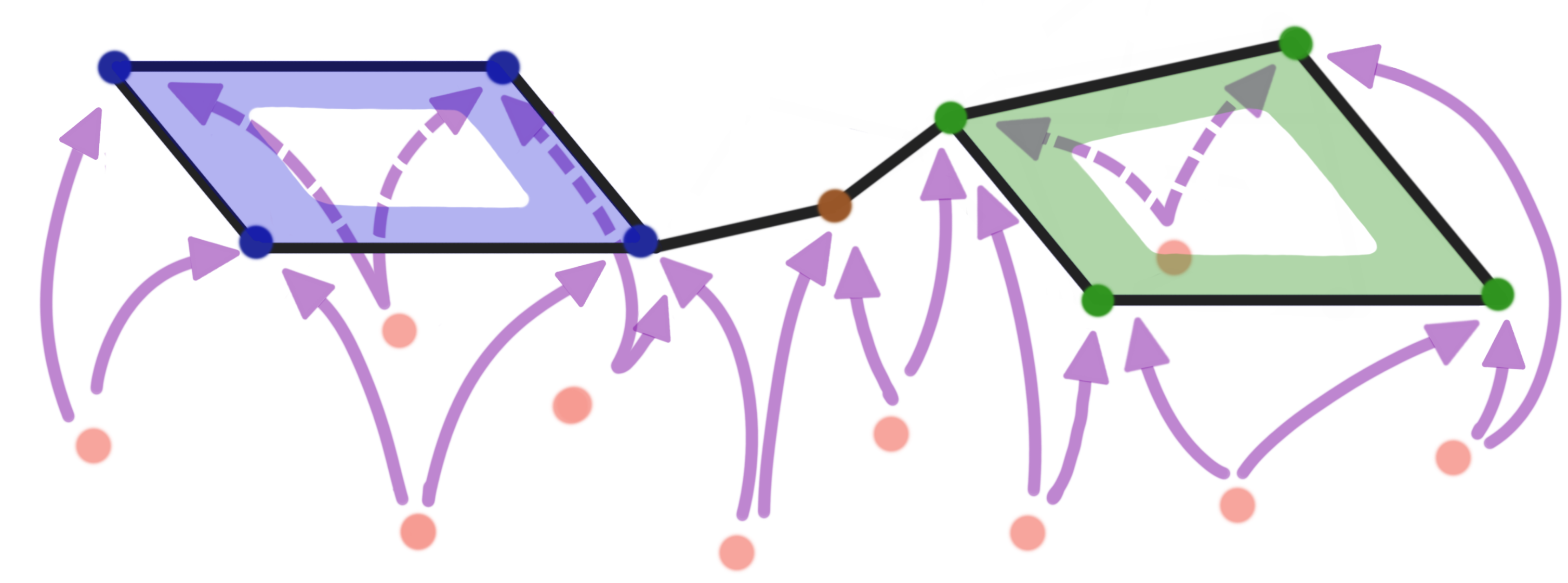}
    \caption{Example 1: Multivariate time series dependence network with two cycles pattern as defined in Equation \ref{eq:two_cycles_model}.}
    \label{fig:time_series_visualization_dependence_1}
\end{figure}

\begin{figure}[b]
    \centering
    \includegraphics[width=.7\linewidth]{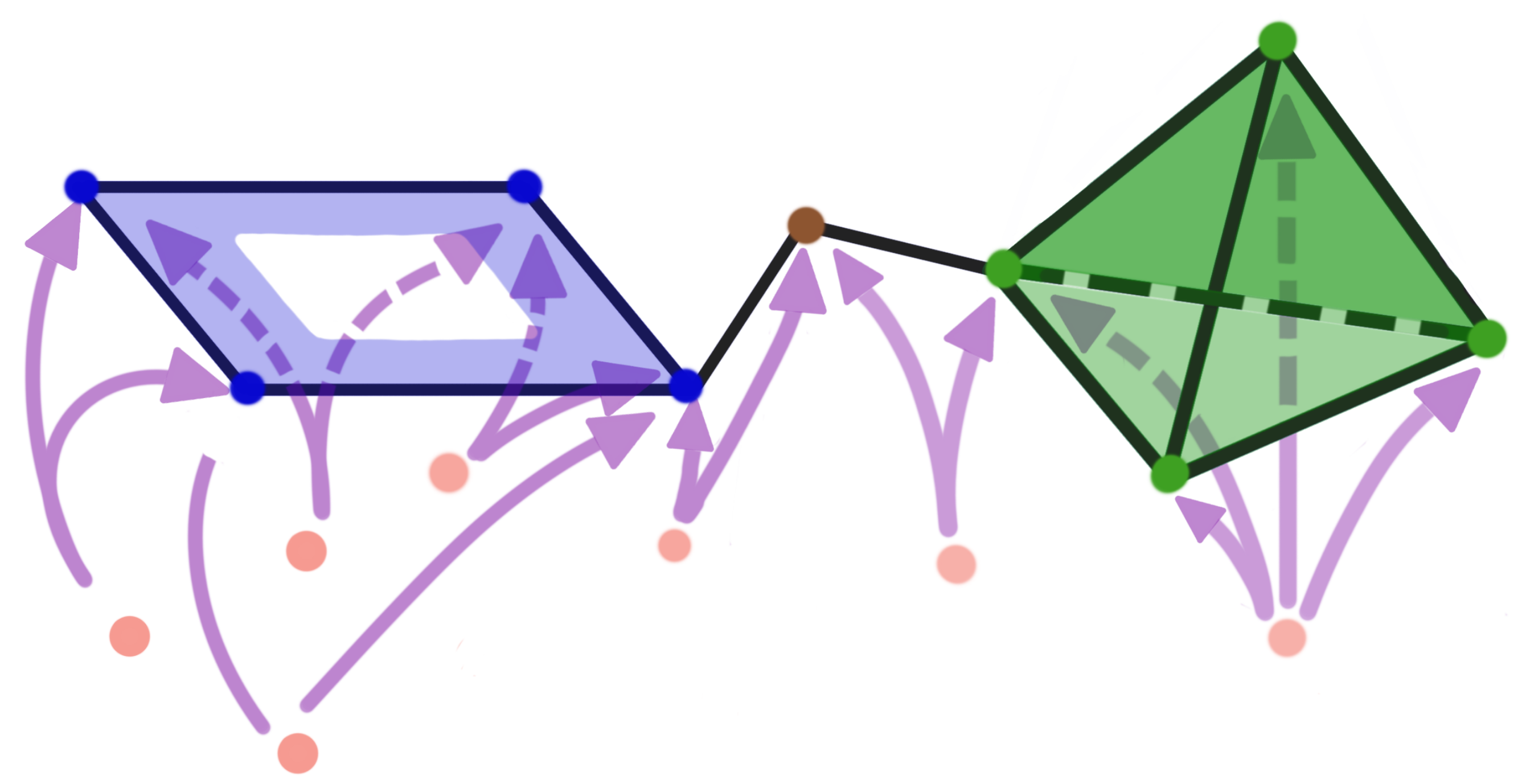}
    \caption{Example 2: Multivariate time series dependence network with a cycle and a 4-clique pattern as defined in Equation \ref{eq:one_cycles_model}.}
    \label{fig:time_series_visualization_dependence_2}
\end{figure}

\begin{figure}[b]
    \centering
    \includegraphics[width=\linewidth]{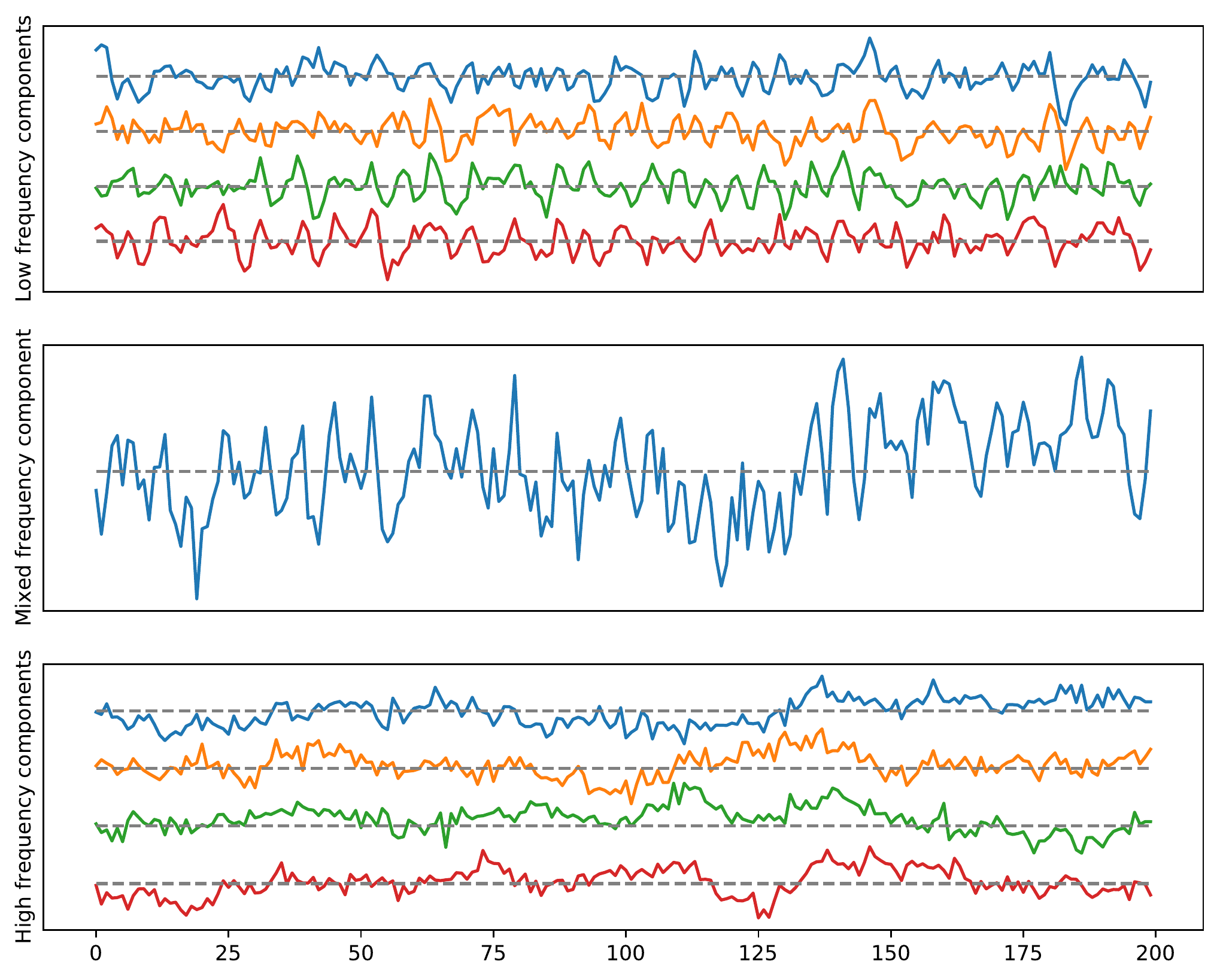}
    \caption{Example 1: A multivariate time series with two loops dependence pattern. High-frequency in the top (cycle), low-frequency in the bottom (cycle).}
    \label{fig:time_series_visualization_1}
\end{figure}

\begin{figure}[b]
    \centering
    \includegraphics[width=\linewidth]{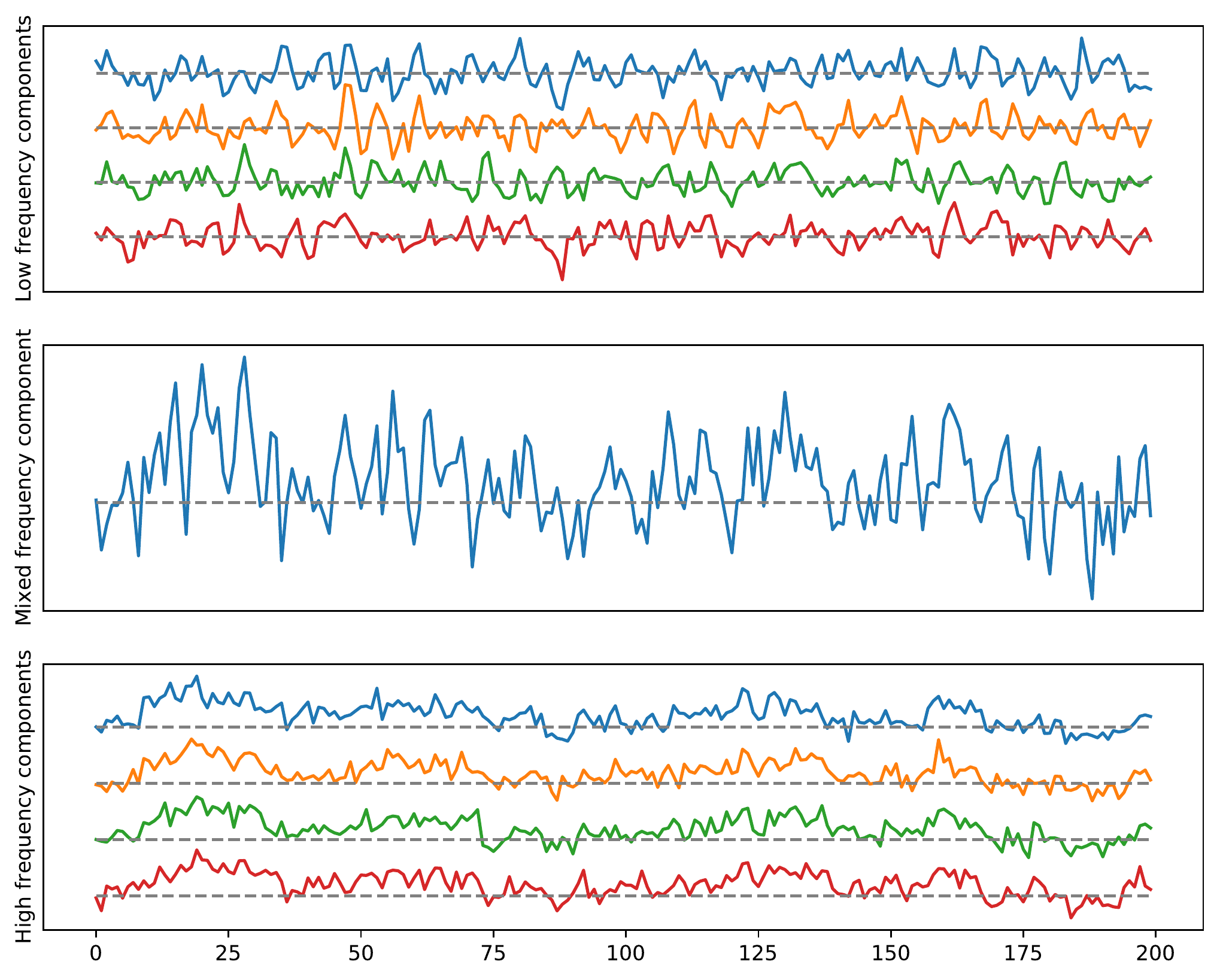}
    \caption{Example 2: A multivariate time series with one loop and a 4-clique dependence pattern. High-frequency in the top (cycle), low-frequency in the bottom (4-clique).}
    \label{fig:time_series_visualization_2}
\end{figure}

\noindent {\bf Examples.} The goal here is to illustrate the previous idea by considering multivariate stationary time series data with a given cyclic dependence network (cyclic frequency specific communities), see Figure \ref{fig:time_series_visualization_dependence_1} and \ref{fig:time_series_visualization_dependence_2}. However, the advantage of applying TDA to the weighted network as explained above is to detect the presence of such topological features. The simulated time series can be visualized in Figure \ref{fig:time_series_visualization_1} and \ref{fig:time_series_visualization_2}.

\begin{figure}[b]
    \centering
    \includegraphics[width=\linewidth]{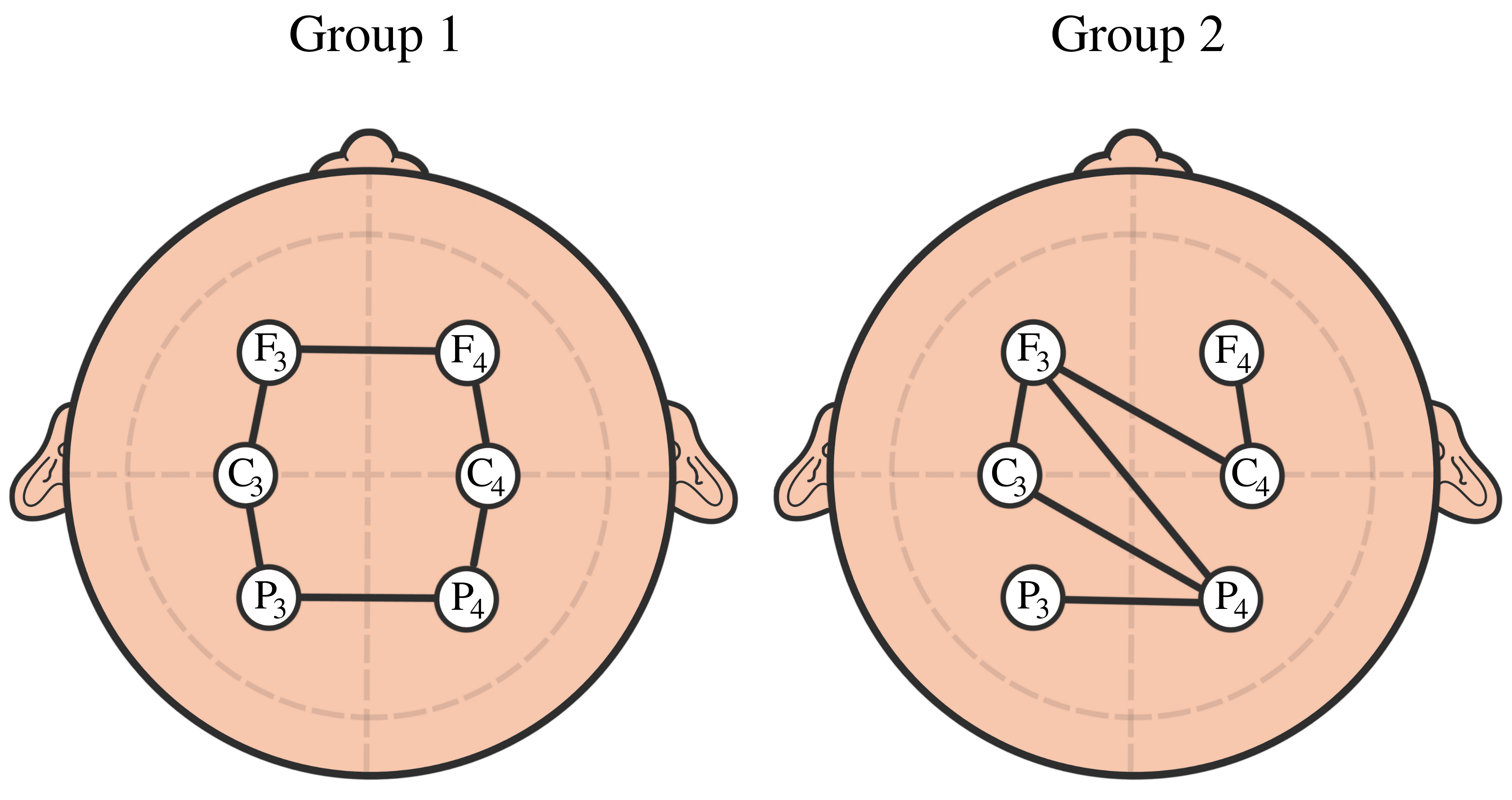}
    \caption{Example of cyclic brain connectivity vs random connectivity.}
    \label{fig:cycle_vs_random}
\end{figure}

It is very common in brain signals to exhibit communities or cyclic structures, as seen in \cite{DYNAMIC_COMUNITY_STRUCTURE} and \cite{BIC_NET} or in simulations in Figure \ref{fig:cycle_vs_random}. There are various reasons that could explain the presence of such pattern in brain networks. For instance, the brain network could be organized in such a way to increase the efficiency of information transfer or minimize the energy consumption. Also, the brain connectivity network could be altered due to some brain damage, e.g., Alzheimer's disease could impair some brain regions that could result in the creation of cycles/holes or voids. In general, we can imagine the following scenarios: 
\begin{itemize}
    \item Groups of neurons firing together (presence of clusters)
    \item Groups of neurons share the some latent processes (potential cycles)
\end{itemize}


To generate the previous multivariate time series (in Figure \ref{fig:time_series_visualization_1} and \ref{fig:time_series_visualization_2}) with both dependence patterns, we use the following approach:

\noindent {\bf Example 1:} The goal of this example is to show how to generate a time series with the dependence pattern presented in Figure \ref{fig:time_series_visualization_dependence_1}. Let $Y(t) = [Y_1(t), \hdots, Y_9(t)]'$ be the observed time series, $Z(t) = [Z_1(t), \hdots, Z_8(t)]'$ the latent AR(2) processes (as in Figure \ref{fig:AR2_time_series_and_spectrum}) and $\epsilon(t) = [\epsilon_1(t), \hdots, \epsilon_9(t)]'$ be the iid Gaussian innovations. Then we can simulate $Y(t)$ as follows:
\begin{align}
    A &= \frac{1}{2}\begin{pmatrix}
    1 & 1 & 0 & 0 & 0 & 0 & 0 & 0\\
    0 & 1 & 1 & 0 & 0 & 0 & 0 & 0\\
    0 & 0 & 1 & 1 & 0 & 0 & 0 & 0\\
    1 & 0 & 0 & 1 & 0 & 0 & 0 & 0\\
    0 & 0 & 0 & 1 & 1 & 0 & 0 & 0\\
    0 & 0 & 0 & 0 & 1 & 1 & 0 & 0\\
    0 & 0 & 0 & 0 & 0 & 1 & 1 & 0\\
    0 & 0 & 0 & 0 & 0 & 0 & 1 & 1\\
    0 & 0 & 0 & 0 & 1 & 0 & 0 & 1\\
    \end{pmatrix}, \\ 
    Y(t) &=  A Z(t) + \epsilon(t). \label{eq:two_cycles_model}
\end{align}
where $\epsilon(t)$ are iid Gaussian white noise with covariance $\sigma_\epsilon = I_9$.
\noindent {\bf Example 2:} The goal of this example is to show how to generate a time series with the dependence pattern presented in Figure \ref{fig:time_series_visualization_dependence_2}. Let $Y(t) = [Y_1(t), \hdots, Y_9(t)]'$ be the observed time series, $Z(t) = [Z_1(t), \hdots, Z_5(t)]'$  be another set of independent latent AR(2) processes (as in Figure \ref{fig:AR2_time_series_and_spectrum}). Then using the following matrix, we can generate the second dependence structure in $Y(t)$.
\begin{align}
    A &= \frac{1}{2}\begin{pmatrix}
    1 & 1 & 0 & 0 & 0\\
    0 & 1 & 1 & 0 & 0\\
    0 & 0 & 1 & 1 & 0\\
    1 & 0 & 0 & 1 & 0\\
    0 & 0 & 0 & 1 & 1\\
    0 & 0 & 0 & 0 & 2\\
    0 & 0 & 0 & 0 & 2\\
    0 & 0 & 0 & 0 & 2\\
    0 & 0 & 0 & 0 & 2\\
    \end{pmatrix}, \\ 
    Y(t) &=  A Z(t) + \epsilon(t) \label{eq:one_cycles_model}
\end{align}
where $\epsilon(t)$ are iid Gaussian white noise with covariance $\sigma_\epsilon = I_9$. 
In order to build the Vietoris-Rips filtration, a dependence-based distance function needs to be defined between the various components of the time series. For instance a decreasing function of any relevant dependence measure could be useful. Therefore, based on the dependence network, a distance matrix can be used to build the persistence homology. First, define the Fourier coefficients and the smoothed periodogram as follows:
\begin{align}
    d(\omega_k) &= \frac{1}{\sqrt{T}} \sum_{t=1}^T X(t)\exp{(-i2\pi t \omega_k)}  \label{Eq:fourier_coef} \\
    \widehat{f}(\omega_k) &= \sum_\omega k_h(\omega - \omega_k) d(\omega_k){d(\omega_k)}^* \label{Eq:smoothed_periodogram}
\end{align} 
where $k_h(\omega - \omega_k)$ is a smoothing kernel centered around $\omega_k$ and $h$ is the bandwidth parameter. Second, define the dependence-based distance function to be a decreasing function (e.g., $\mathcal{G}(x) = 1-x$) of coherence:
\begin{align}
    \mathcal{C}\Big(X_{i}(.), X_{j}(.), \omega\Big) &= \frac{|f_{i, j}(\omega)|^2}{f_{i, i}(\omega)f_{j, j}(\omega)} \in [0, 1]. \label{eq:coherence} \\
    \mathcal{D}\Big(X_{i}(.), X_{j}(.), \omega\Big) &= \mathcal{G}\Big(\mathcal{C}(X_{i}(.), X_{j}(.), \omega)\Big). \label{eq:dependence_based_distance}
\end{align}
Coherence at some pre-specified frequency band is the squared cross-correlation between a pair of filtered signals (whose power are each concentrated at the specific band), see \cite{ONBAO_BELLEGEM}. Another way to estimate coherence is via the maximal cross-correlation-squared of the bandpass filtered signals, see \cite{SPECTRAL_DEPENDENCE}.

\noindent {\bf Example 3:} The goal of this example is to provide a simple model that can explain/distinguish the observed cyclic structure from the random one in some data sets, see Figure \ref{fig:cycle_vs_random}. Assume, we observe six time series components that share common cyclic latent independent process copies $Z_i(t)$ for group 1:

\begin{align*}
    y_1(t) &= Z_6(t) + Z_1(t) + c \epsilon_1(t) \\
    y_2(t) &= Z_1(t) + Z_2(t) + c \epsilon_2(t) \\
    y_3(t) &= Z_2(t) + Z_3(t) + c \epsilon_3(t) \\
    y_4(t) &= Z_3(t) + Z_4(t) + c \epsilon_4(t) \\
    y_5(t) &= Z_4(t) + Z_5(t) + c \epsilon_5(t) \\
    y_6(t) &= Z_5(t) + Z_6(t) + c \epsilon_6(t)
\end{align*}
Assume, we observe five time series components that share random common latent independent process copies $Z_i(t)$ for group 2:
\begin{align*}
    y_1(t) &= Z_6(t) + Z_3(t) + Z_4(t) + c \epsilon_1(t) \\
    y_2(t) &= Z_6(t) + Z_2(t) + c \epsilon_2(t) \\
    y_3(t) &= Z_1(t) + c \epsilon_3(t) \\
    y_4(t) &= Z_1(t) + Z_2(t) + Z_3(t) + c \epsilon_4(t) \\
    y_5(t) &= Z_4(t) + Z_5(t) + c \epsilon_5(t) \\
    y_6(t) &= Z_5(t) + c \epsilon_6(t)
\end{align*}
when the parameter $c=0$ (vanishing noise) the coherence between the observed time series is maximal: Coh$(y_i, y_j, \omega) = 1, \forall \omega \in \Omega_\alpha$, and as the parameter $c$ increases the coherence between the observed time series drops to reach zero when $c=\infty$. In this way, the signal to noise ratio (SNR=$\frac{Var(Signal)}{Var(Noise)}$) controls the coherence between the components and therefore it controls the distance between the components. This simple example aims to demonstrate a potential explanation of the mechanism behind the appearance of topological features in the brain dependence network.
\begin{figure}[b]
    \centering
    \includegraphics[width=\linewidth]{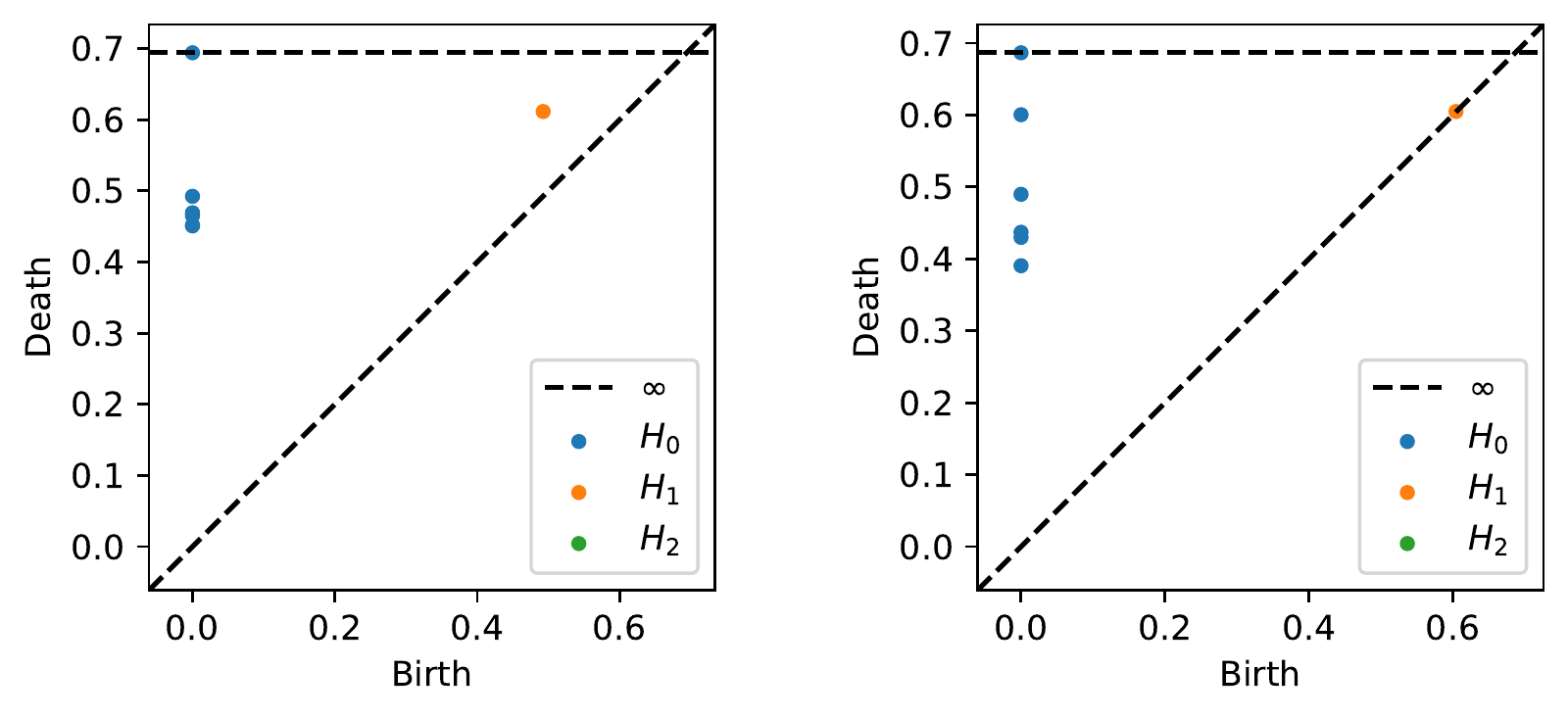}
    \caption{Persistence diagram based on the previous example displaying cyclic or group 1 (LEFT) and random brain connectivity or group 2 (RIGHT).}
    \label{fig:PD_cycle_vs_random}
\end{figure}
After generating the time series data from both models as described previously, we estimated the coherence matrix based on the smoothed periodogram (rectangular window) for the 100-200Hz frequency band. Based on this we construct the distance matrix and apply TDA to get the PD in Figure \ref{fig:PD_cycle_vs_random}. We can clearly see that the subplot on the left display 1-dimensional features (orange dot far from the diagonal), whereas the subplot on the right does not.

\begin{figure}
    \centering
    \includegraphics[width=\linewidth]{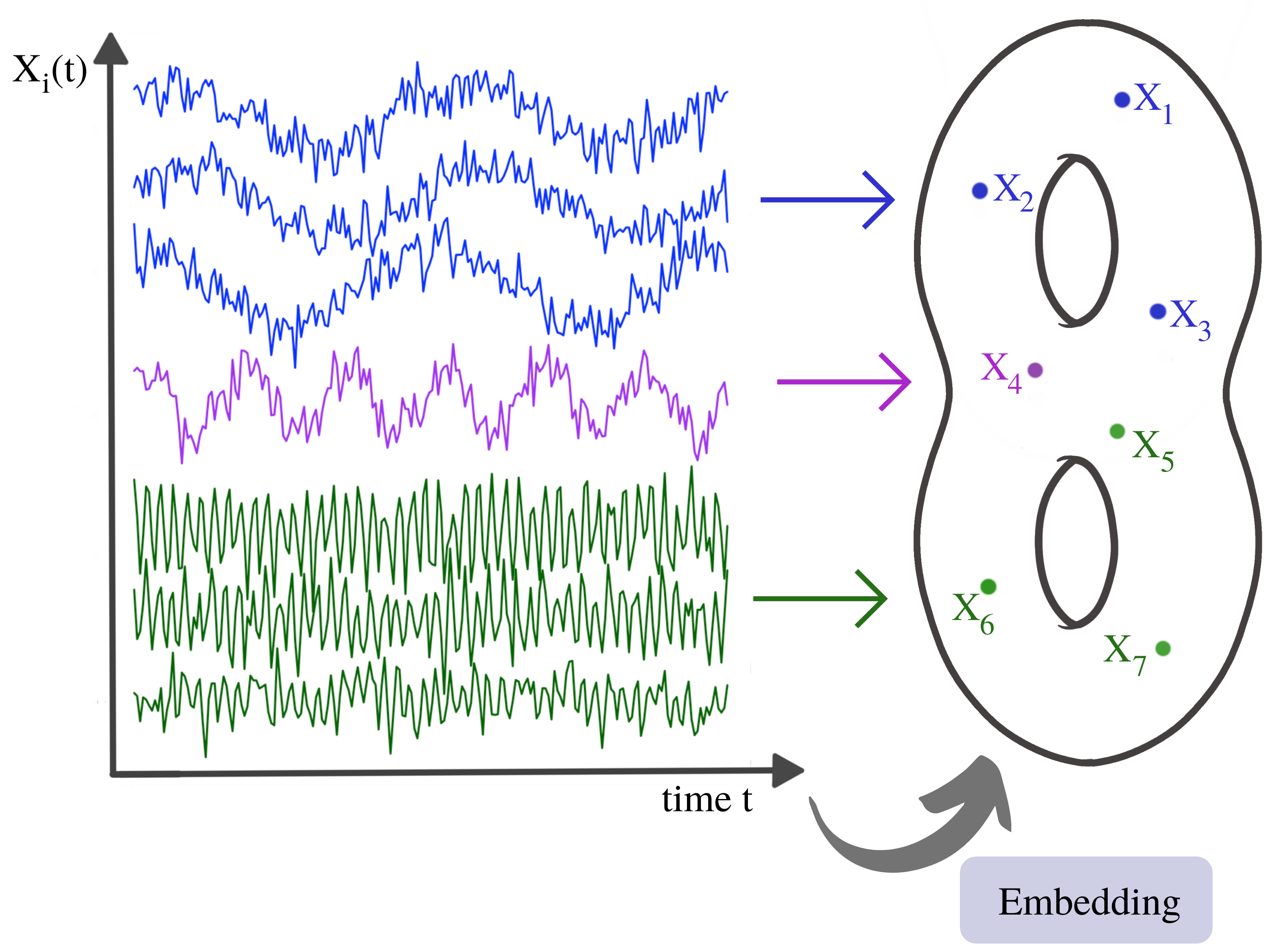}
    \caption{Illustration of a time series dependence embedding in some abstract dependence space.}
    \label{fig:time_series_embedding_illustration_3}
\end{figure}

To summarize, what we are trying to learn from the previous examples is the underlying topological pattern of the dependence space. We can try to visualize this idea using Figure \ref{fig:time_series_embedding_illustration_3}.

\subsection{TDA vs Graph theoretical modeling of brain connectivity}

The human brain is organized both structurally and functionally as well into complex networks. This complex structure allows both the segregation and integration of information processing. The classical approach to analyze brain functional connectivity consists of using tools from the science of networks, which often involves the use of graph theoretical summaries such as modularity, efficiency and betweenness, see \cite{GRAPH_MODELING_BRAIN_CONNECTIVITY}. Graph theoretical models witnessed an important success in modeling complex brain networks, as it is described in \cite{GRAPH_MODELING_COMPLEX_NETWORK} and \cite{GRAPH_MODELING_COMPLEX_NETWORK_STRUCT_FUNCT}. The above mentioned graph theoretical summaries aim to characterize the topological properties of the network being studied, and hence can distinguish between small-world, scale-free and random networks. Indeed, such graph theoretical models display important features that are of particular interest in the study of brain activity. For example, random networks (a network where the edges are selected randomly) usually have a low clustering coefficient (low measure of the degree to which nodes in a graph tend to cluster together) and a low characteristic path length (low average distance between pairs of nodes), on the other hand, regular networks have a
high clustering coefficient but a high characteristic path length. However, small-world models have high clustering (higher than random graphs) and low characteristic path (roughly the logarithm of the size of the graph). Furthermore, scale-free networks can have even smaller characteristic path length and potentially smaller clustering coefficient than that of small-world networks. Such topological properties may have a direct impact on brain activity such as the robustness to brain injury or efficiency of information transfer between far apart brain regions (variable cost of brain integration). See \cite{SMALL_WORLD_NETWORKS} for an overview of small-world networks and their potential applications and properties and \cite{SCALE_FREE_NETWORKS} for an overview of of the emergence of scale-free networks in random networks. Such models and graph summaries have been extensively used to study the impact of diseases on the topology of brain connectivity, see \cite{GRAPH_MODELING_AD_BRAIN} and \cite{GRAPH_MODELING_HUMAN_BRAIN}. 

\begin{figure}
    \centering
    \includegraphics[width=\linewidth]{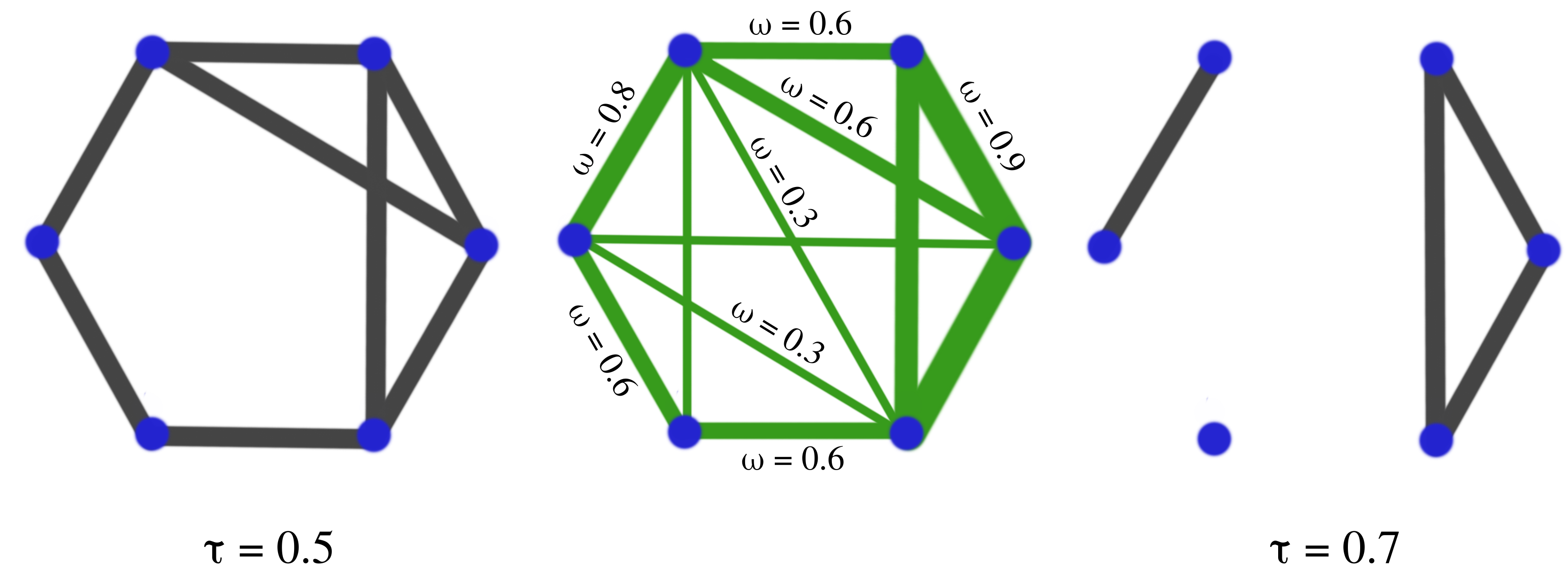}
    \caption{The thresholding step consists in comparing the weights of the original network (CENTER) with some given threshold $\tau$. If the weight of the edge is larger than the threshold, one edge is created in the new network, if the weight is smaller, no edge is added. If the selected threshold is low ($\tau=0.5$), the resulting network is dense (LEFT), if the threshold is too high ($\tau=0.7$) the resulting network is sparse (RIGHT). The problem becomes: How to select the threshold $\tau$ so we balance the loss of information with sparsity?}
    \label{fig:thresholding}
\end{figure}

\begin{figure}
    \centering
    \includegraphics[width=.7\linewidth]{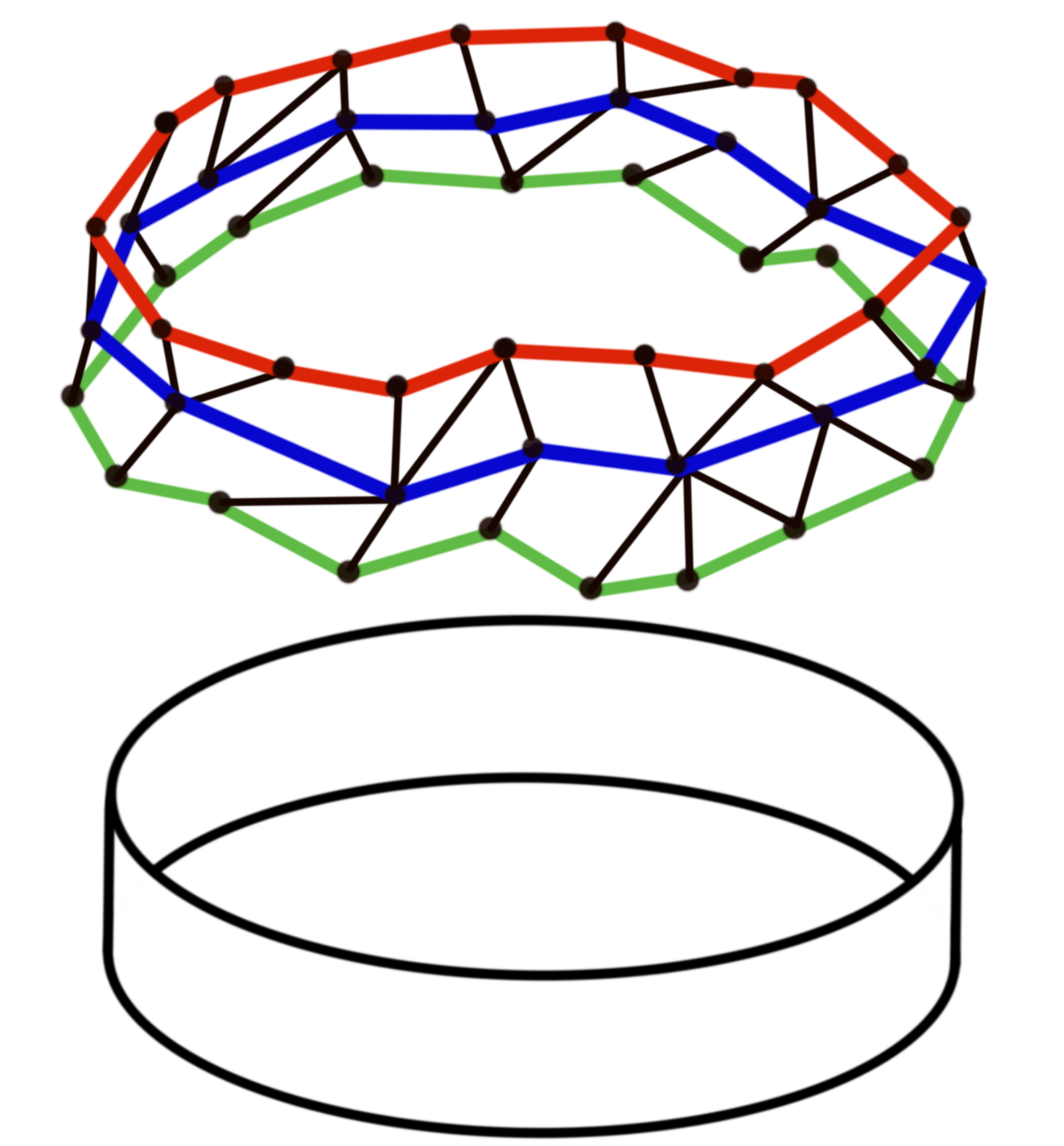}
    \caption{Network with a cyclic feature. Original network (TOP), topological structure being represented (BOTTOM).}
    \label{fig:topological_cycle}
\end{figure}

The goal of such approach is to characterize the topological properties of the brain network. Although such summaries provide interesting and valuable insights into the topology of brain networks, they nevertheless suffer some limitations. Indeed, such summaries cannot capture all the topological information contained in the network, such as the presence of holes and voids, see Figure \ref{fig:topological_cycle}. Furthermore, such summaries cannot be applied directly to a weighed connectivity network. Very often, a thresholding step is necessary, which can be a serious limitation because it can result in an important loss of information if the threshold is not selected properly, as seen in Figure \ref{fig:thresholding}. In this regard, applying TDA to brain connectivity could provide complementary information on the topology of the brain functional network, since TDA consider all potential threshold values.

There are many advantages to using the persistence homology techniques. Topological data analysis is designed to study the topological features (geometry and spatial organization) of networks. Classical approaches describe some topological properties of the network. However, it remains difficult to detect/assess the topological patterns present in general, as seen in Figure \ref{fig:topological_cycle}. A graph theoretical algorithm will count as three different cycles, the cycles around the same hole (green, blue and red). However, TDA can detect exactly only one large hole/topological feature because it uses the concept of a simplicial complexes. Furthermore, an algorithm that clusters the network nodes (modularity analysis) needs a parameter choice, whereas the TDA techniques provides overall answers regarding the network topology without parameter tuning.

\section{Data application and permutation testing}
\label{sec:data_application_testing}

The purpose of this section is to compare the differences in the topological features of the brain connectivity networks of young individuals with Attention Deficit Hyperactivity Disorder (ADHD) and healthy control groups, see \cite{ADHD_DATA}. Specifically, the impact of ADHD on the connected components (0-dimensional homology) and the network cyclic information (1-dimensional homology).

\begin{figure}
    \centering
    \includegraphics[width=.5\linewidth]{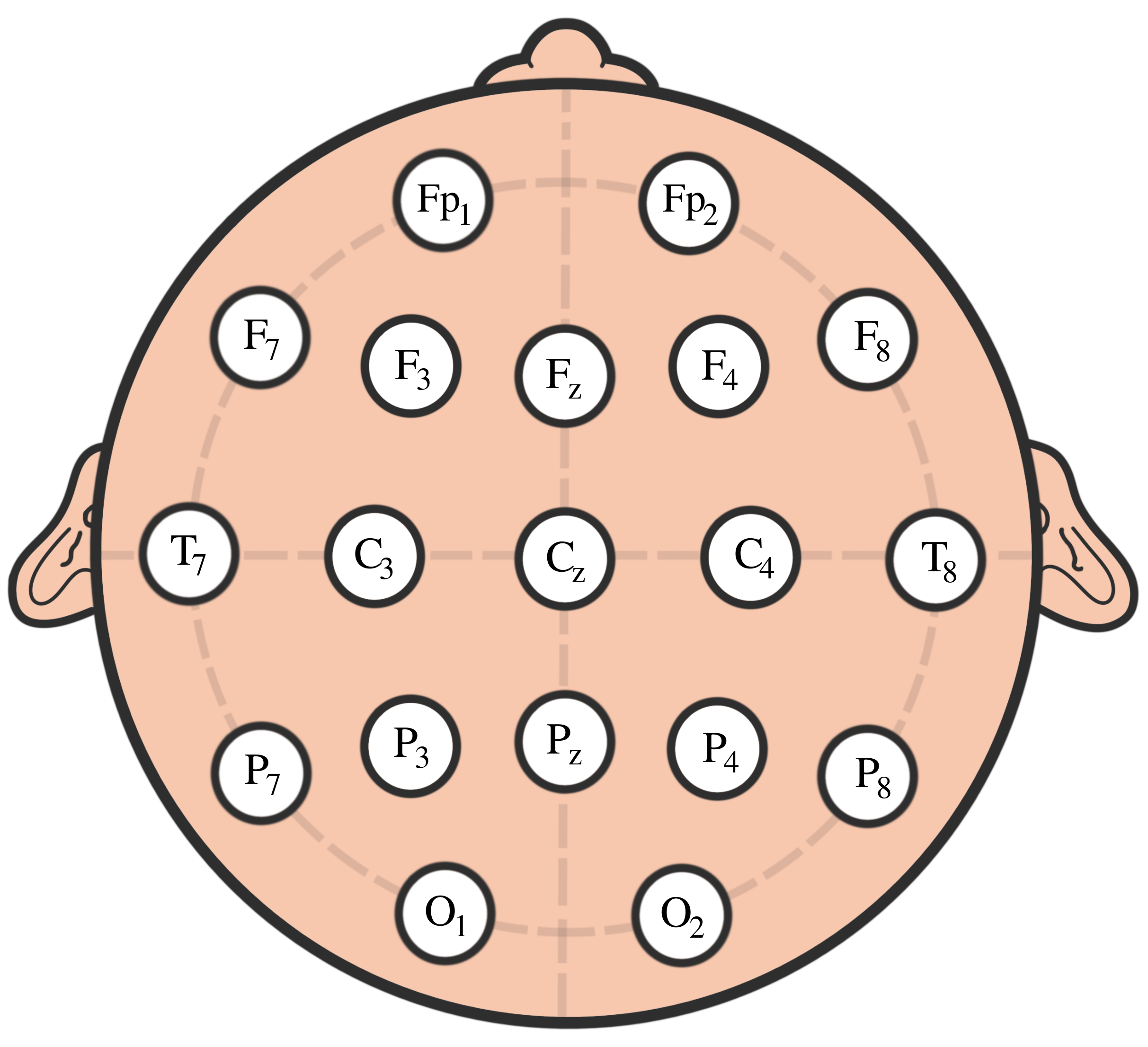}
    \caption{Scalp EEG with 10-20 standard layout.}
    \label{fig:scalp_eeg}
\end{figure}

The participants in this study were 61 children with ADHD and 60 healthy controls aged between 7 - 12 years old. The ADHD children were diagnosed by an experienced psychiatrist, and have taken Ritalin for up to 6 months. None of the children in the control group had a history of psychiatric disorders, epilepsy, or any report of high-risk behaviors. EEG signals were recorded based on 10-20 standard by 19 channels at a sampling frequency of 128 Hz, see Figure \ref{fig:scalp_eeg}.

Since one of the deficits of ADHD children is visual attention, the EEG recording protocol was based on a visual attention task. The children were shown a series of cartoon characters and were asked to count them. To ensure a continuous stimulus throughout the signal recording, each image was displayed immediately and uninterrupted after the child's response. Hence, the duration of EEG recording throughout this cognitive visual task was based on the child's response speed. Only 51 subjects with ADHD were kept with 53 healthy controls. The PREP pipeline was used to preprocess the data (removing the effect of the electrical line; removing artifacts due to eye movements eye blinks or muscular movements; detecting/removing/repairing bad quality channels; filtering non-relevant signal components and finally re-referencing the signal to improve topographical localization).

\begin{figure}[b]
    \centering
    \includegraphics[width=\linewidth]{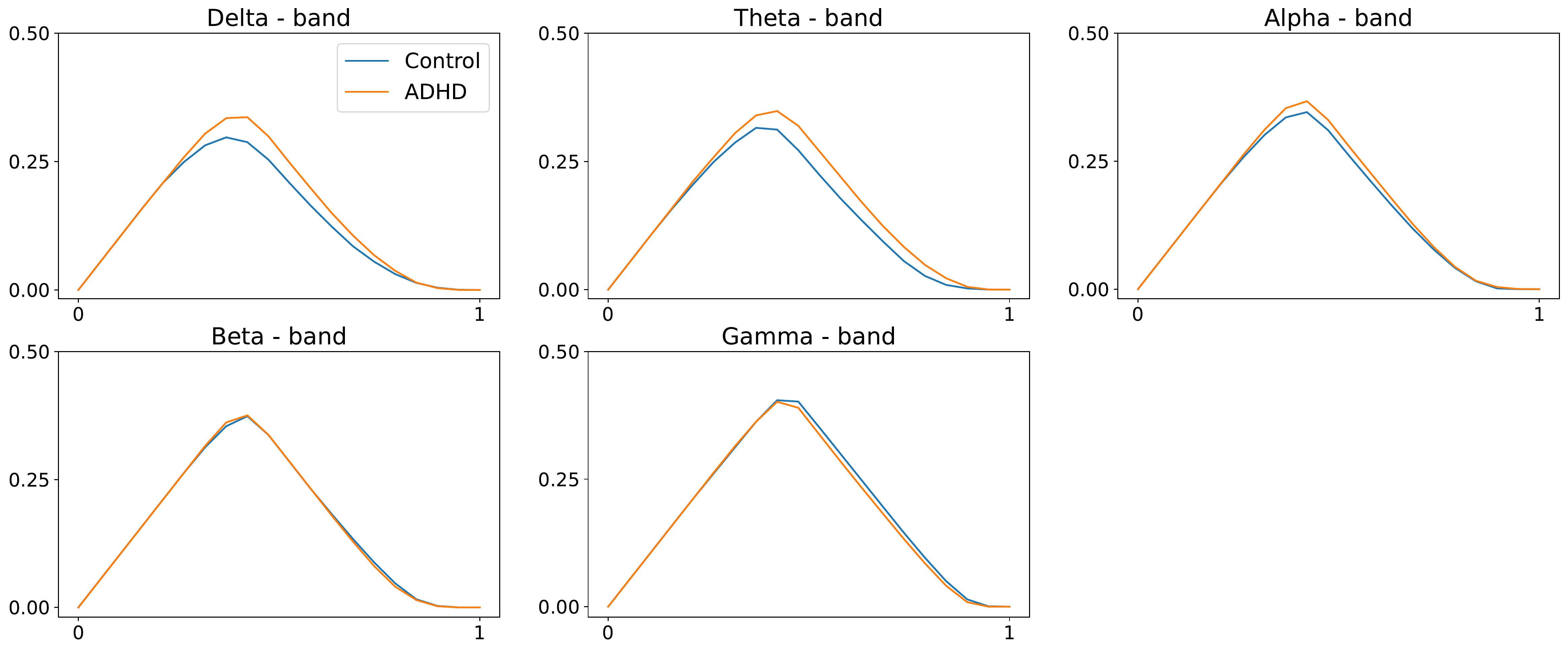}
    \caption{Population average persistence landscapes for the 0-dimensional homology group for ADHD (orange) and healthy control (blue) groups, at various frequency bands. High frequency bands do not seem to display any differences between the two groups. These plots suggest that both groups have a similar structure at the connected components level.}
    \label{fig:persistance_landscapes_0_Homology}
\end{figure}

\begin{figure}[b]
    \centering
    \includegraphics[width=\linewidth]{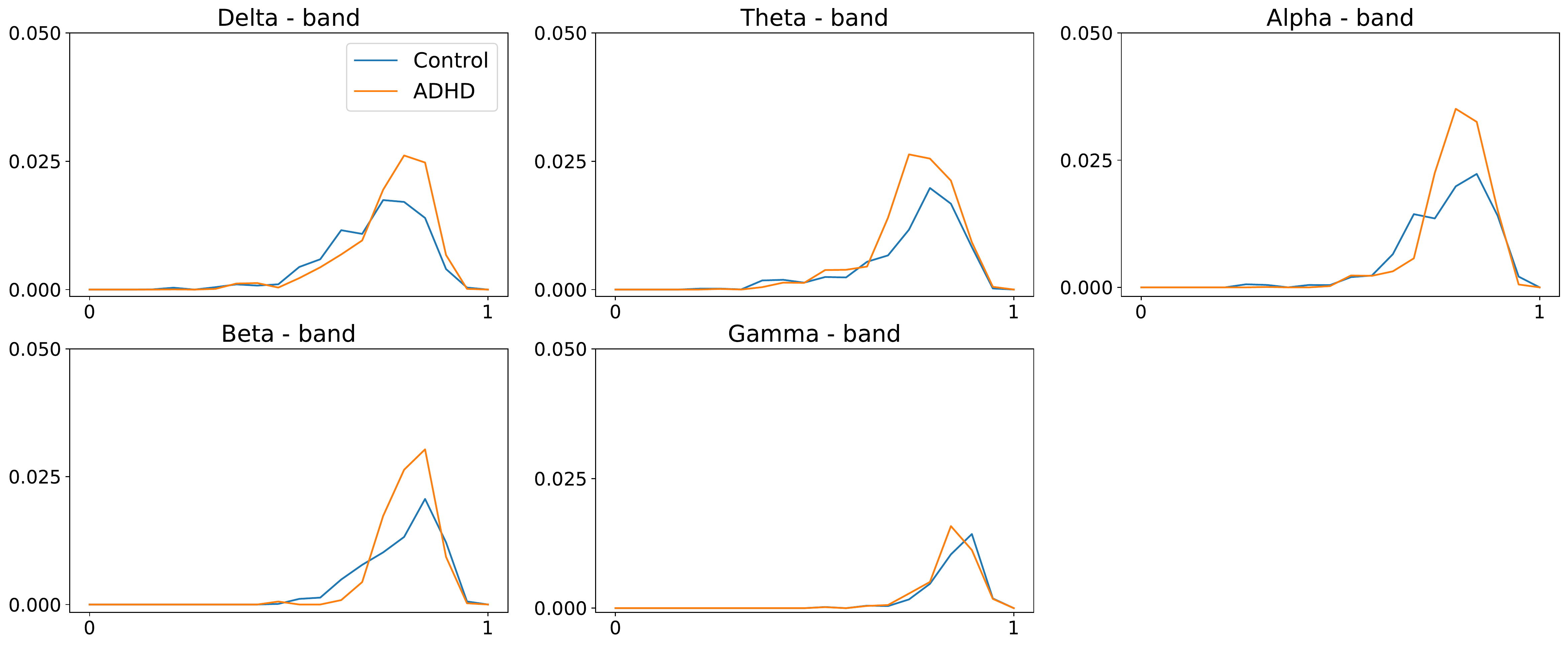}
    \caption{Population average persistence landscapes for the 1-dimensional homology group for ADHD (orange) and healthy control (blue) groups, at various frequency bands. Middle frequency bands seem to display differences between the two groups. This suggests that the ADHD group seem to have more cycles/holes in their dependence network.}
    \label{fig:persistance_landscapes_1_Homology}
\end{figure}

We compute the average coherence matrix for different frequency bands in Hertz (delta = (0.5-4Hz), theta = (4, 8Hz), alpha = (8-12Hz), beta = (12-30Hz) and gamma = (30-50Hz)) and use it to build the distance function ($D = 1 - C$), as defined in Equation \ref{eq:dependence_based_distance}. For every frequency band, we build the persistence landscapes for both groups: ADHD and healthy controls, see Figure \ref{fig:persistance_landscapes_0_Homology} and \ref{fig:persistance_landscapes_1_Homology}.

% \begin{figure}[b]
%     \centering
%     \includegraphics[width=\linewidth]{images/control_vs_ADHD_2_Homology_landscapes.pdf}
%     \caption{Population average persistence landscape (two dimensional homology group) for both ADHD and healthy control groups, at different frequency bands.}
%     \label{fig:persistance_landscapes_2_Homology}
% \end{figure}

In both Figures \ref{fig:persistance_landscapes_0_Homology} and \ref{fig:persistance_landscapes_1_Homology}, we see the group level differences in the persistence landscapes. For the zero-dimensional homology, the differences seem to be at the $delta$, $theta$ and $alpha$-frequency bands. In addition, the one-dimensional homology displays group differences at all frequency bands except the $gamma$-frequency bands. On contrast, the two-dimensional homology group seems to display differences between the two groups that are too small in magnitude (one order of magnitude smaller than the previous ones!).

Now, assume that we are interested in testing for differences in the topology of the brain dependence network between the two group (ADHD and healthy control). Let $H_0$ be the null hypothesis: \'There is no impact of ADHD on the brain connectivity network\'. In order to test such hypothesis, we may carry a permutation test based on the norm of the discrepancy persistence landscapes at some given homology dimension and for specific frequency band $\Omega$:
\begin{align}
    T^\Omega_k = \int_{\Omega} \Big|\Big| \overline{\lambda}^{(1)}_k(\omega) - \overline{\lambda}^{(2)}_k(\omega) \Big|\Big|_2  d\omega
\end{align}
In order to make a decision to reject $H_0$, we need to compare the observed test statistic with a threshold obtained from the reference distribution of the test statistic under $H_0$. We use a permutation approach to derive this empirical distribution under the null hypothesis, as it was done in \cite{PL_FIRST}, \cite{EXPERIMENTAL_RANDOMIZATION_1} or \cite{TDA_MORSE_EEG}. A formal framework for testing between two groups in topological data analysis is presented in \cite{HYPOTHESIS_TESTING_TDA}, with an extension to three groups in \cite{HYPOTHESIS_TESTING_TDA_EXTENSION}. Practical examples on nonparametric permutation tests at an acceptable level can be found in \cite{PERMUTATION_TESTING_NEUROIMAGING}. Refer to \cite{EXPERIMENTAL_RANDOMIZATION_1} and \cite{EXPERIMENTAL_RANDOMIZATION_2} for more examples regarding permutation and randomization tests in functional brain imaging and connectivity. Therefore, following the permutation approach we propose the following procedure:
\begin{enumerate}
    \item Compute the sample test statistic from the original PLs: $\lambda_1^{(1)}, \hdots, \lambda_{n_1}^{(1)}$ and $\lambda_1^{(2)}, \hdots, \lambda_{n_2}^{(2)}$.
    \item Permute the ADHD and healthy control group labels to get $\lambda_1^{(1*)}, \hdots, \lambda_{n_1}^{(1*)}$ and $\lambda_1^{(2*)}, \hdots, \lambda_{n_2}^{(2*)}$
    \item Compute the sample discrepancy from the permuted PLs: $$T^{\Omega*} = \int_{\Omega} \big|\big| \overline{\lambda}^{(1*)}_k(\omega) - \overline{\lambda}^{(2*)}_k(\omega) \big|\big|_2 d\omega$$
    \item Repeat steps 2 to 3, $B$ times
    \item Compute the threshold $\tau$ as the ($1-\alpha$)-quantile of the empirical distribution of test statistic $\widehat{F}_{T}$
\end{enumerate}
After applying the above to our data we obtain the following reference distribution for the zero- and one-dimensional homology persistence landscapes.

\begin{figure}
    \centering
    \includegraphics[width=\linewidth]{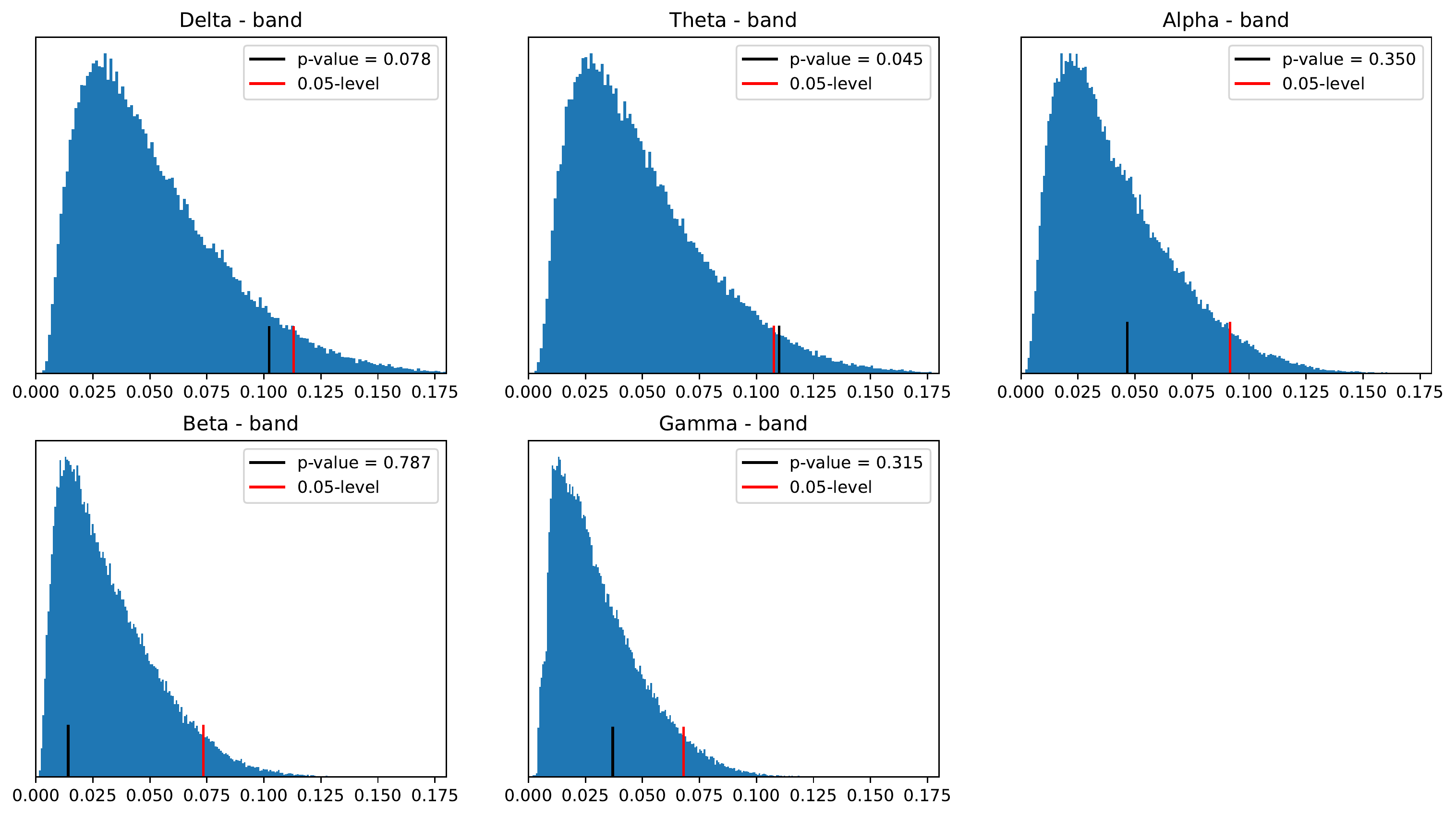}
    \caption{Reference distribution for testing for group level differences between ADHD and healthy control persistence diagrams, based on B=100000 permutations. Zero-dimensional homology group.}
    \label{fig:permutation_reference distributions_0_Homology}
\end{figure}

\begin{figure}
    \centering
    \includegraphics[width=\linewidth]{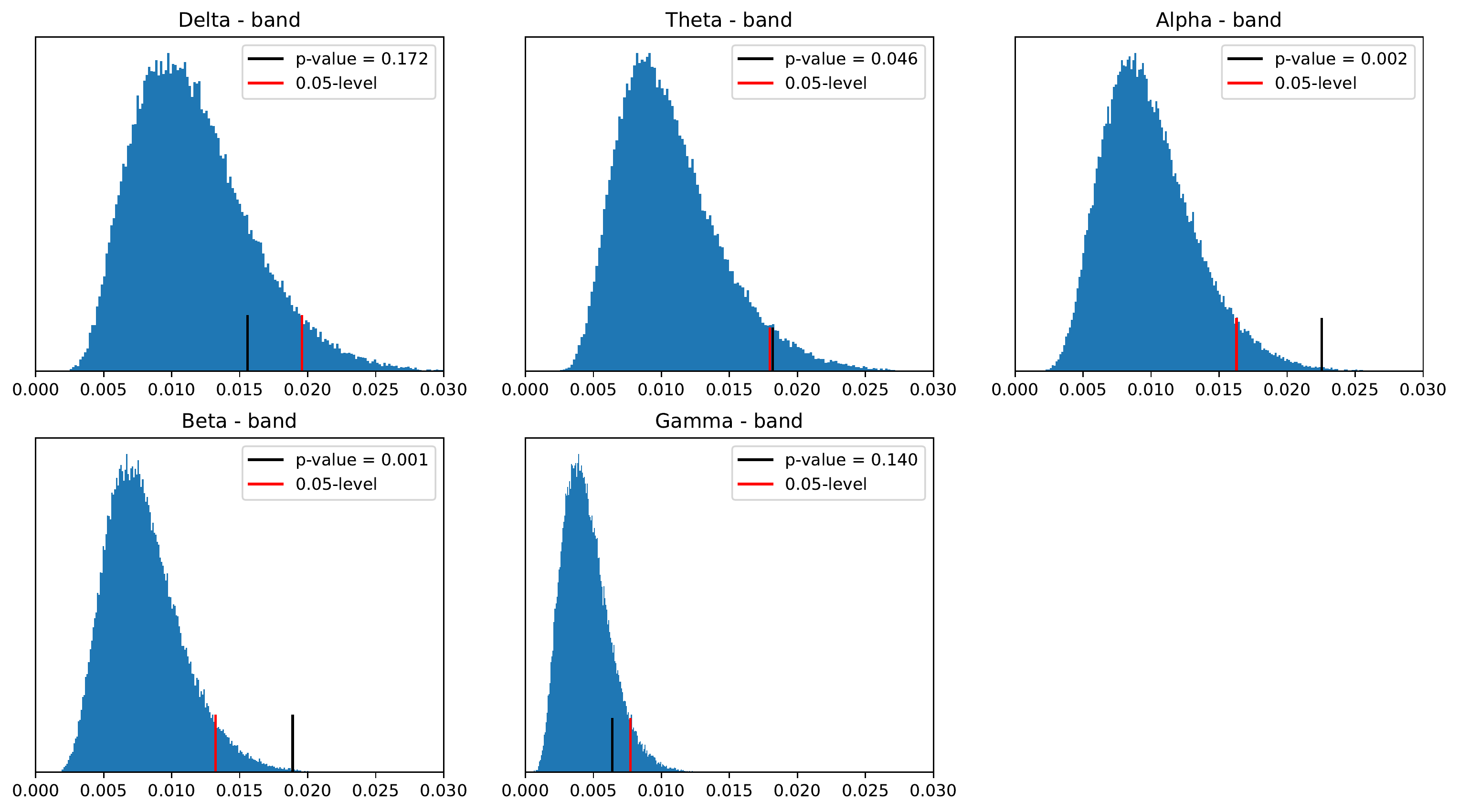}
    \caption{Reference distribution for testing for group level differences between ADHD and healthy control persistence diagrams, based on B=100000 permutations. One-dimensional homology group.}
    \label{fig:permutation_reference distributions_1_Homology}
\end{figure}

Despite the differences between the PLs of the two populations in zero and one-dimensional homology groups, only the differences in the $alpha$- and $beta$-frequency bands seem to be significant for the one-dimensional homology group, as shown in Figures \ref{fig:persistance_landscapes_0_Homology} and \ref{fig:persistance_landscapes_1_Homology}.

\vspace{-2mm}
\section{Open problems}
\vspace{-1mm}

As the field of topological data analysis keeps advancing and developing, new and challenging problems continue to emerge. We discuss briefly three open problems that may be of interest to readers with an interest in topological data analysis applied to brain networks.

The study of brain signals shows that the brain dependence networks may display between group discrepancy as well as within group variability. Historically, linear mixed-effects models (LMEM) have been proposed to to analyze data with fixed effects (average persistence landscape) and random effects (variance of the persistence landscape), e.g., $y = X\beta + Uz + \epsilon$). Is it possible to develop such a model, which can be applied to detect group level differences (fixed effect, i.e., differences in $\beta$) in the topological structure of the network via the estimated persistence landscapes as well as within group variations of the topological structure (random effects, i.e., differences in $z$).

Brain dependence networks can be constructed based on various dependence measures. When correlation or coherence are used to measure dependence between brain channels the resulting dependence network is a non-oriented graph. In contrast, when more complex models of dependence (e.g., flow of information) are used to model the dependencies between brain channels, such as partial directed coherence, the resulting network is an oriented one. This results in a non-symmetric distance function which is a problem for the application of TDA. A potential approach to extend the use of TDA to oriented networks is to use the matrix decomposition $A = A^s + A^n$, where $A^s=\frac{1}{2}(A + A^{'})$ and $A^n=\frac{1}{2}(A - A^{'})$.

The classical application of TDA results in a global analysis of the network. Therefore, it is impossible to state where the topological features are located in the network. Therefore, it natural to wonder if TDA can be applied locally? to "local sub-networks"? Can we think of TDA in a hierarchical terms? 
Similarly, sometimes a transient change in connectivity can be observed in brain networks (e.g., localized behavior in time that leads to task-specific functional brain connectivity). How can we use TDA to study such evolutionary or transient events?

\vspace{-2mm}
\section{Conclusion}
\vspace{-1mm}
\label{sec:conclusion}

Historically, brain network analysis relied on graph theoretical measures such as clustering coefficients, betweeness centrality, and average shortest path length to study the topology. Although such an approach revealed some interesting facts about the brain in the past, it does not give us the full picture of the network's geometry. In contrast, TDA has begun to be used to analyze brain network topological data from a persistent homology perspective. This enables a summary of all the scales without having to use arbitrary thresholds.

The purpose of the paper was to provide pedagogical introduction to topological data analysis within a multivariate spectral analysis of time series data. This approach has the advantage of combining the power of TDA with spectral analysis. Which will allow practitioners to characterize the commonalities and differences in the shape of brain connectivity networks across different groups for frequency-specific neural oscillations. 

We demonstrated the advantages of using the Rips-Vietoris filtration over the Morse one. We gave a pedagogical review of persistent homology using the Vietoris-Rips filtration over a cloud of points. We discussed how the time delay embedding could be pertinent and showed its limits when the initial assumptions of the Takens's theorem are not satisfied. Finally, we recommended to apply TDA to the connectivity network, as this could capture the rich information contained in the brain signals dependence structure as this was shown in the data application.

Indeed, the application of TDA to the connectivity networks of ADHD vs healthy control individuals shows significant discrepancy between their respective PLs at the alpha- and beta-frequency bands for the one-homology group. This suggests that the ADHD condition affects the cyclic structure of the brain connectivity network more than the connected components.

%%%%%%%%%%%%%%%%%%%%%%%%%%%%%%%%%%%%%%%%%%%%%%
%% Single Appendix:                         %%
%%%%%%%%%%%%%%%%%%%%%%%%%%%%%%%%%%%%%%%%%%%%%%
%\begin{appendix}
%\section*{???}%% if no title is needed, leave empty \section*{}.
%\end{appendix}
%%%%%%%%%%%%%%%%%%%%%%%%%%%%%%%%%%%%%%%%%%%%%%
%% Multiple Appendixes:                     %%
%%%%%%%%%%%%%%%%%%%%%%%%%%%%%%%%%%%%%%%%%%%%%%
%\begin{appendix}
%\section{???}
%
%\section{???}
%
%\end{appendix}

%%%%%%%%%%%%%%%%%%%%%%%%%%%%%%%%%%%%%%%%%%%%%%
%% Support information, if any,             %%
%% should be provided in the                %%
%% Acknowledgements section.                %%
%%%%%%%%%%%%%%%%%%%%%%%%%%%%%%%%%%%%%%%%%%%%%%
\begin{acks}[Acknowledgments]
The authors would like to thank Sarah Aracid (KAUST and University of the Philippines) for her invaluable help with the figures and artwork.
\end{acks}
%%%%%%%%%%%%%%%%%%%%%%%%%%%%%%%%%%%%%%%%%%%%%%
%% Funding information, if any,             %%
%% should be provided in the                %%
%% funding section.                         %%
%%%%%%%%%%%%%%%%%%%%%%%%%%%%%%%%%%%%%%%%%%%%%%
\begin{funding}
The authors were supported by King Abdullah University of Science and Technology (KAUST) Research Fund.
\end{funding}

%%%%%%%%%%%%%%%%%%%%%%%%%%%%%%%%%%%%%%%%%%%%%%
%% Supplementary Material, including data   %%
%% sets and code, should be provided in     %%
%% {supplement} environment with title      %%
%% and short description. It cannot be      %%
%% available exclusively as external link.  %%
%% All Supplementary Material must be       %%
%% available to the reader on Project       %%
%% Euclid with the published article.       %%
%%%%%%%%%%%%%%%%%%%%%%%%%%%%%%%%%%%%%%%%%%%%%%
%\begin{supplement}
%\stitle{???}
%\sdescription{???.}
%\end{supplement}

%%%%%%%%%%%%%%%%%%%%%%%%%%%%%%%%%%%%%%%%%%%%%%%%%%%%%%%%%%%%%
%%                  The Bibliography                       %%
%%                                                         %%
%%  imsart-???.bst  will be used to                        %%
%%  create a .BBL file for submission.                     %%
%%                                                         %%
%%  Note that the displayed Bibliography will not          %%
%%  necessarily be rendered by Latex exactly as specified  %%
%%  in the online Instructions for Authors.                %%
%%                                                         %%
%%  MR numbers will be added by VTeX.                      %%
%%                                                         %%
%%  Use \cite{...} to cite references in text.             %%
%%                                                         %%
%%%%%%%%%%%%%%%%%%%%%%%%%%%%%%%%%%%%%%%%%%%%%%%%%%%%%%%%%%%%%

% \clearpage
% \newpage
\bibliographystyle{imsart-number} 
% (imsart-number.bst or imsart-nameyear.bst)
\bibliography{references}

\begin{thebibliography}{58}
% BibTex style file: imsart-number.bst, 2017-11-03
% Default style options (sort=1,type=number).
% Used options (sort=1,type=number).

\bibitem{TDA_MORSE_RANDOM_FIELDS}
\begin{barticle}[author]
\bauthor{\bsnm{Adler},~\bfnm{Robert}\binits{R.}},
  \bauthor{\bsnm{Bobrowski},~\bfnm{Omer}\binits{O.}},
  \bauthor{\bsnm{Borman},~\bfnm{Matthew}\binits{M.}},
  \bauthor{\bsnm{Subag},~\bfnm{Eliran}\binits{E.}} \AND
  \bauthor{\bsnm{Weinberger},~\bfnm{Shmuel}\binits{S.}}
(\byear{2010}).
\btitle{Persistent Homology for Random Fields and Complexes}.
\bjournal{Borrowing Strength: Theory Powering Applications}
\bvolume{6}
\bpages{124-143}.
\bdoi{10.1214/10-IMSCOLL609}
\end{barticle}
\endbibitem

\bibitem{WASSERSTEIN_BOTELLNECK}
\begin{barticle}[author]
\bauthor{\bsnm{Agami},~\bfnm{Sarit}\binits{S.}}
(\byear{2021}).
\btitle{Comparison of persistence diagrams}.
\bjournal{Communications in Statistics - Simulation and Computation}
\bvolume{0}
\bpages{1-14}.
\bdoi{10.1080/03610918.2021.1894335}
\end{barticle}
\endbibitem

\bibitem{LECTURES_MORSE_HOMOLOGY}
\begin{bbook}[author]
\bauthor{\bsnm{Banyaga},~\bfnm{Augustin}\binits{A.}} \AND
  \bauthor{\bsnm{Hurtubise},~\bfnm{David}\binits{D.}}
(\byear{2004}).
\btitle{Lectures on Morse Homology}
\bvolume{29}.
\bdoi{10.1007/978-1-4020-2696-6}
\end{bbook}
\endbibitem

\bibitem{SCALE_FREE_NETWORKS}
\begin{barticle}[author]
\bauthor{\bsnm{Barabasi},~\bfnm{Albert-Laszlo}\binits{A.-L.}} \AND
  \bauthor{\bsnm{Albert},~\bfnm{Reka}\binits{R.}}
(\byear{1999}).
\btitle{Emergence of Scaling in Random Networks}.
\bjournal{Science}
\bvolume{286}
\bpages{509-512}.
\bdoi{10.1126/science.286.5439.509}
\end{barticle}
\endbibitem

\bibitem{GRAPH_MODELING_HUMAN_BRAIN}
\begin{barticle}[author]
\bauthor{\bsnm{Bassett},~\bfnm{Danielle~S.}\binits{D.~S.}} \AND
  \bauthor{\bsnm{Bullmore},~\bfnm{Edward~T.}\binits{E.~T.}}
(\byear{2009}).
\btitle{Human brain networks in health and disease}.
\bjournal{Current Opinion in Neurology}
\bvolume{22}
\bpages{340–347}.
\bdoi{10.1097/WCO.0b013e32832d93dd}
\end{barticle}
\endbibitem

\bibitem{TDA_BRAIN_ARTERY}
\begin{barticle}[author]
\bauthor{\bsnm{Bendich},~\bfnm{Paul}\binits{P.}},
  \bauthor{\bsnm{Marron},~\bfnm{J.~S.}\binits{J.~S.}},
  \bauthor{\bsnm{Miller},~\bfnm{Ezra}\binits{E.}},
  \bauthor{\bsnm{Pieloch},~\bfnm{Alex}\binits{A.}} \AND
  \bauthor{\bsnm{Skwerer},~\bfnm{Sean}\binits{S.}}
(\byear{2016}).
\btitle{{Persistent homology analysis of brain artery trees}}.
\bjournal{The Annals of Applied Statistics}
\bvolume{10}
\bpages{198-218}.
\bdoi{10.1214/15-AOAS886}
\end{barticle}
\endbibitem

\bibitem{NERVE_THEOREM}
\begin{barticle}[author]
\bauthor{\bsnm{Bj{\"o}rner},~\bfnm{Anders}\binits{A.}}
(\byear{1995}).
\btitle{Topological methods}.
\bjournal{Handbook of combinatorics}
\bvolume{2}
\bpages{1819--1872}.
\end{barticle}
\endbibitem

\bibitem{OPTIMAL_THRESHOLDING}
\begin{barticle}[author]
\bauthor{\bsnm{Bordier},~\bfnm{Cécile}\binits{C.}},
  \bauthor{\bsnm{Nicolini},~\bfnm{Carlo}\binits{C.}} \AND
  \bauthor{\bsnm{Bifone},~\bfnm{Angelo}\binits{A.}}
(\byear{2017}).
\btitle{Graph Analysis and Modularity of Brain Functional Connectivity
  Networks: Searching for the Optimal Threshold}.
\bjournal{Frontiers in Neuroscience}
\bvolume{11}.
\bdoi{10.3389/fnins.2017.00441}
\end{barticle}
\endbibitem

\bibitem{PL_FIRST}
\begin{barticle}[author]
\bauthor{\bsnm{Bubenik},~\bfnm{Peter}\binits{P.}}
(\byear{2015}).
\btitle{Statistical Topological Data Analysis Using Persistence Landscapes}.
\bjournal{Journal of Machine Learning Research}
\bvolume{16}
\bpages{77–102}.
\bdoi{10.5555/2789272.2789275}
\end{barticle}
\endbibitem

\bibitem{GRAPH_MODELING_COMPLEX_NETWORK_STRUCT_FUNCT}
\begin{barticle}[author]
\bauthor{\bsnm{Bullmore},~\bfnm{Edward}\binits{E.}} \AND
  \bauthor{\bsnm{Sporns},~\bfnm{Olaf}\binits{O.}}
(\byear{2009}).
\btitle{Complex brain networks: Graph theoretical analysis of structural and
  functional systems}.
\bjournal{Nature reviews. Neuroscience}
\bvolume{10}
\bpages{186-98}.
\bdoi{10.1038/nrn2575}
\end{barticle}
\endbibitem

\bibitem{BRAIN_ACTIVITY_NETWORK_REST}
\begin{barticle}[author]
\bauthor{\bsnm{Cabral},~\bfnm{Joana}\binits{J.}},
  \bauthor{\bsnm{Kringelbach},~\bfnm{Morten~L.}\binits{M.~L.}} \AND
  \bauthor{\bsnm{Deco},~\bfnm{Gustavo}\binits{G.}}
(\byear{2014}).
\btitle{Exploring the network dynamics underlying brain activity during rest}.
\bjournal{Progress in Neurobiology}
\bvolume{114}
\bpages{102-131}.
\bdoi{https://doi.org/10.1016/j.pneurobio.2013.12.005}
\end{barticle}
\endbibitem

\bibitem{TDA_BRAIN_PROS_AND_CONS}
\begin{barticle}[author]
\bauthor{\bsnm{Caputi},~\bfnm{Luigi}\binits{L.}},
  \bauthor{\bsnm{Pidnebesna},~\bfnm{Anna}\binits{A.}} \AND
  \bauthor{\bsnm{Hlinka},~\bfnm{Jaroslav}\binits{J.}}
(\byear{2021}).
\btitle{Promises and pitfalls of topological data analysis for brain
  connectivity analysis}.
\bjournal{NeuroImage}
\bvolume{238}
\bpages{118-245}.
\bdoi{https://doi.org/10.1016/j.neuroimage.2021.118245}
\end{barticle}
\endbibitem

\bibitem{TDA_GUNNAR}
\begin{barticle}[author]
\bauthor{\bsnm{Carlsson},~\bfnm{Gunnar}\binits{G.}}
(\byear{2009}).
\btitle{Topology and Data}.
\bjournal{Bulletin of the American Mathematical Society}
\bvolume{46}
\bpages{255-308}.
\bdoi{10.1090/S0273-0979-09-01249-X}
\end{barticle}
\endbibitem

\bibitem{HYPOTHESIS_TESTING_TDA_EXTENSION}
\begin{barticle}[author]
\bauthor{\bsnm{Cericola},~\bfnm{Christopher}\binits{C.}},
  \bauthor{\bsnm{Johnson},~\bfnm{Inga~Jo}\binits{I.~J.}},
  \bauthor{\bsnm{Kiers},~\bfnm{Joshua}\binits{J.}},
  \bauthor{\bsnm{Krock},~\bfnm{Mitchell}\binits{M.}},
  \bauthor{\bsnm{Purdy},~\bfnm{Jordan}\binits{J.}} \AND
  \bauthor{\bsnm{Torrence},~\bfnm{Johanna}\binits{J.}}
(\byear{2018}).
\btitle{Extending hypothesis testing with persistent homology to three or more
  groups}.
\bjournal{Involve: A Journal of Mathematics}
\bvolume{11}
\bpages{27-51}.
\bdoi{10.2140/involve.2018.11.27}
\end{barticle}
\endbibitem

\bibitem{TDA_MORSE_CORTICAL_DATA}
\begin{barticle}[author]
\bauthor{\bsnm{Chung},~\bfnm{Moo~K.}\binits{M.~K.}},
  \bauthor{\bsnm{Bubenik},~\bfnm{Peter}\binits{P.}} \AND
  \bauthor{\bsnm{Kim},~\bfnm{Peter~T.}\binits{P.~T.}}
(\byear{2009}).
\btitle{Persistence diagrams of cortical surface data}.
\bjournal{Information processing in medical imaging}
\bvolume{21}
\bpages{403-414}.
\bdoi{10.1007/978-3-642-02498-6_32}
\end{barticle}
\endbibitem

\bibitem{PD_STABILITY}
\begin{barticle}[author]
\bauthor{\bsnm{Cohen-Steiner},~\bfnm{David}\binits{D.}},
  \bauthor{\bsnm{Edelsbrunner},~\bfnm{Herbert}\binits{H.}} \AND
  \bauthor{\bsnm{Harer},~\bfnm{John}\binits{J.}}
(\byear{2007}).
\btitle{Stability of persistence diagrams}.
\bjournal{Discrete and computational geometry}
\bvolume{37}
\bpages{103-120}.
\bdoi{doi.org/10.1007/s00454-006-1276-5}
\end{barticle}
\endbibitem

\bibitem{EDELSBRUNNER_HARER}
\begin{barticle}[author]
\bauthor{\bsnm{Edelsbrunner},~\bfnm{Herbert}\binits{H.}} \AND
  \bauthor{\bsnm{Harer},~\bfnm{John}\binits{J.}}
(\byear{2008}).
\btitle{Persistent homology—a survey}.
\bjournal{Discrete and Computational Geometry}
\bvolume{453}
\bpages{257-282}.
\bdoi{10.1090/conm/453/08802}
\end{barticle}
\endbibitem

\bibitem{TDA_EDELSBRUNNER}
\begin{barticle}[author]
\bauthor{\bsnm{Edelsbrunner},~\bfnm{Herbert}\binits{H.}},
  \bauthor{\bsnm{Letscher},~\bfnm{David}\binits{D.}} \AND
  \bauthor{\bsnm{Zomorodian},~\bfnm{Afra}\binits{A.}}
(\byear{2002}).
\btitle{Topological Persistence and Simplification}.
\bvolume{28}
\bpages{511–533}.
\bdoi{doi.org/10.1007/s00454-002-2885-2}
\end{barticle}
\endbibitem

\bibitem{EXPERIMENTAL_RANDOMIZATION_2}
\begin{barticle}[author]
\bauthor{\bsnm{Fiecas},~\bfnm{Mark}\binits{M.}},
  \bauthor{\bsnm{Ombao},~\bfnm{Hernando}\binits{H.}},
  \bauthor{\bsnm{Linkletter},~\bfnm{Crystal}\binits{C.}},
  \bauthor{\bsnm{Thompson},~\bfnm{Wesley}\binits{W.}} \AND
  \bauthor{\bsnm{Sanes},~\bfnm{Jerome}\binits{J.}}
(\byear{2010}).
\btitle{Functional connectivity: Shrinkage estimation and randomization test}.
\bjournal{NeuroImage}
\bvolume{49}
\bpages{3005-3014}.
\bdoi{https://doi.org/10.1016/j.neuroimage.2009.12.022}
\end{barticle}
\endbibitem

\bibitem{BARCODES}
\begin{barticle}[author]
\bauthor{\bsnm{Ghrist},~\bfnm{Robert}\binits{R.}}
(\byear{2008}).
\btitle{Barcodes: The persistent topology of data}.
\bjournal{Bulletin of the American Mathematical Society}
\bvolume{45}
\bpages{61-75}.
\bdoi{10.1090/S0273-0979-07-01191-3}
\end{barticle}
\endbibitem

\bibitem{TDA_GUIDEA}
\begin{barticle}[author]
\bauthor{\bsnm{Gidea},~\bfnm{Marian}\binits{M.}} \AND
  \bauthor{\bsnm{Katz},~\bfnm{Yuri}\binits{Y.}}
(\byear{2018}).
\btitle{Topological data analysis of financial time series: Landscapes of
  crashes}.
\bjournal{Physica A: Statistical Mechanics and its Applications}
\bvolume{491}
\bpages{820-834}.
\bdoi{10.1016/j.physa.2017.09.028}
\end{barticle}
\endbibitem

\bibitem{TDA_FINANCIAL_TSA}
\begin{barticle}[author]
\bauthor{\bsnm{Gidea},~\bfnm{Marian}\binits{M.}} \AND
  \bauthor{\bsnm{Katz},~\bfnm{Yuri~A.}\binits{Y.~A.}}
(\byear{2018}).
\btitle{Topological Data Analysis of Financial Time Series: Landscapes of
  Crashes}.
\bjournal{Physica A-statistical Mechanics and Its Applications}
\bvolume{491}
\bpages{820-834}.
\bdoi{10.1016/j.physa.2017.09.028}
\end{barticle}
\endbibitem

\bibitem{AR2_MIXTURE}
\begin{barticle}[author]
\bauthor{\bsnm{Granados-Garcia},~\bfnm{Guillermo}\binits{G.}},
  \bauthor{\bsnm{Fiecas},~\bfnm{Mark}\binits{M.}},
  \bauthor{\bsnm{Shahbaba},~\bfnm{Babak}\binits{B.}},
  \bauthor{\bsnm{Fortin},~\bfnm{Norbert}\binits{N.}} \AND
  \bauthor{\bsnm{Ombao},~\bfnm{Hernando}\binits{H.}}
(\byear{2021}).
\btitle{Modeling Brain Waves as a Mixture of Latent Processes}.
\bdoi{https://arxiv.org/pdf/2102.11971.pdf}
\end{barticle}
\endbibitem

\bibitem{EXTREMAL_CON}
\begin{barticle}[author]
\bauthor{\bsnm{Guerrero},~\bfnm{Matheus}\binits{M.}},
  \bauthor{\bsnm{Huser},~\bfnm{Raphaël}\binits{R.}} \AND
  \bauthor{\bsnm{Ombao},~\bfnm{Hernando}\binits{H.}}
(\byear{2021}).
\btitle{Conex-Connect: Learning Patterns in Extremal Brain Connectivity From
  Multi-Channel EEG Data}.
\bdoi{https://arxiv.org/pdf/2101.09352.pdf}
\end{barticle}
\endbibitem

\bibitem{EEG_EXPERIMENT}
\begin{barticle}[author]
\bauthor{\bsnm{Hasenstab},~\bfnm{Kyle}\binits{K.}},
  \bauthor{\bsnm{Sugar},~\bfnm{Catherine}\binits{C.}},
  \bauthor{\bsnm{Telesca},~\bfnm{Donatello}\binits{D.}},
  \bauthor{\bsnm{Mcevoy},~\bfnm{Kevin}\binits{K.}},
  \bauthor{\bsnm{Jeste},~\bfnm{Shafali}\binits{S.}} \AND
  \bauthor{\bsnm{Şentürk},~\bfnm{Damla}\binits{D.}}
(\byear{2015}).
\btitle{Identifying longitudinal trends within EEG experiments}.
\bjournal{Biometrics}
\bvolume{71}
\bpages{1090-1100}.
\bdoi{10.1111/biom.12347}
\end{barticle}
\endbibitem

\bibitem{HAUSMANN_RIPS_FILTRATION}
\begin{bbook}[author]
\bauthor{\bsnm{Hausmann},~\bfnm{Jean-Claude}\binits{J.-C.}}
(\byear{2016}).
\btitle{On the Vietoris-Rips complexes and a Cohomology Theory for metric
  spaces}.
\bpublisher{Princeton University Press}.
\bdoi{doi:10.1515/9781400882588-013}
\end{bbook}
\endbibitem

\bibitem{GRAPH_MODELING_AD_BRAIN}
\begin{barticle}[author]
\bauthor{\bsnm{He},~\bfnm{Yong}\binits{Y.}},
  \bauthor{\bsnm{Chen},~\bfnm{Zhang}\binits{Z.}},
  \bauthor{\bsnm{Gong},~\bfnm{Gaolang}\binits{G.}} \AND
  \bauthor{\bsnm{Evans},~\bfnm{Alan}\binits{A.}}
(\byear{2009}).
\btitle{Neuronal Networks in Alzheimer's Disease}.
\bjournal{The Neuroscientist}
\bvolume{15}
\bpages{333-350}.
\bdoi{10.1177/1073858409334423}
\end{barticle}
\endbibitem

\bibitem{GRAPH_MODELING_BRAIN_CONNECTIVITY}
\begin{barticle}[author]
\bauthor{\bsnm{He},~\bfnm{Yong}\binits{Y.}} \AND
  \bauthor{\bsnm{Evans},~\bfnm{Alan}\binits{A.}}
(\byear{2010}).
\btitle{Graph theoretical modeling of brain connectivity}.
\bjournal{Current opinion in neurology}
\bvolume{23}
\bpages{341-350}.
\bdoi{10.1097/WCO.0b013e32833aa567}
\end{barticle}
\endbibitem

\bibitem{MULTICHANNEL_BRAIN_SIGNALS}
\begin{barticle}[author]
\bauthor{\bsnm{Hu},~\bfnm{Lechuan}\binits{L.}},
  \bauthor{\bsnm{Fortin},~\bfnm{Norbert}\binits{N.}} \AND
  \bauthor{\bsnm{Ombao},~\bfnm{Hernando}\binits{H.}}
(\byear{2017}).
\btitle{Modeling High-Dimensional Multichannel Brain Signals}.
\bjournal{Statistics in Biosciences}
\bvolume{11}.
\bdoi{10.1007/s12561-017-9210-3}
\end{barticle}
\endbibitem

\bibitem{TOPOLOGY_HISTORY}
\begin{barticle}[author]
\bauthor{\bsnm{James},~\bfnm{Ioan~Mackenzie}\binits{I.~M.}}
(\byear{1996}).
\btitle{Reflections on the history of topology}.
\bjournal{Seminario Matematico e Fisico di Milano}
\bvolume{66}
\bpages{87-96}.
\bdoi{10.1007/BF02925355}
\end{barticle}
\endbibitem

\bibitem{PROBLEM_THRESHOLDING_SW_NETWORKS}
\begin{barticle}[author]
\bauthor{\bsnm{Langer},~\bfnm{Nicolas}\binits{N.}},
  \bauthor{\bsnm{Pedroni},~\bfnm{Andreas}\binits{A.}} \AND
  \bauthor{\bsnm{Jäncke},~\bfnm{Lutz}\binits{L.}}
(\byear{2013}).
\btitle{The Problem of Thresholding in Small-World Network Analysis}.
\bjournal{PloS one}
\bvolume{8}
\bpages{e53199}.
\bdoi{10.1371/journal.pone.0053199}
\end{barticle}
\endbibitem

\bibitem{BOOK_FMRI_LAZAR}
\begin{bbook}[author]
\bauthor{\bsnm{Lazar},~\bfnm{Nicole}\binits{N.}}
(\byear{2008}).
\btitle{The Statistical Analysis of Functional MRI Data}.
\bpublisher{Springer}.
\bdoi{10.1007/978-0-387-78191-4}
\end{bbook}
\endbibitem

\bibitem{TDA_BRAIN}
\begin{barticle}[author]
\bauthor{\bsnm{Lee},~\bfnm{Hyekyoung}\binits{H.}},
  \bauthor{\bsnm{Kang},~\bfnm{Hyejin}\binits{H.}},
  \bauthor{\bsnm{Chung},~\bfnm{Moo~K.}\binits{M.~K.}},
  \bauthor{\bsnm{Kim},~\bfnm{Bung-Nyun}\binits{B.-N.}} \AND
  \bauthor{\bsnm{Lee},~\bfnm{Dong~Soo}\binits{D.~S.}}
(\byear{2012}).
\btitle{Persistent Brain Network Homology From the Perspective of Dendrogram}.
\bjournal{IEEE Transactions on Medical Imaging}
\bvolume{31}
\bpages{2267-2277}.
\bdoi{10.1109/TMI.2012.2219590}
\end{barticle}
\endbibitem

\bibitem{FMRI_LINDQUIST}
\begin{barticle}[author]
\bauthor{\bsnm{Lindquist},~\bfnm{Martin~A.}\binits{M.~A.}}
(\byear{2008}).
\btitle{{The Statistical Analysis of fMRI Data}}.
\bjournal{Statistical Science}
\bvolume{23}
\bpages{439-464}.
\bdoi{10.1214/09-STS282}
\end{barticle}
\endbibitem

\bibitem{MERKULOV_ALG_TOP}
\begin{barticle}[author]
\bauthor{\bsnm{Merkulov},~\bfnm{Sergei}\binits{S.}}
(\byear{2003}).
\btitle{Algebraic topology}.
\bjournal{Proceedings of the Edinburgh Mathematical Society}
\bvolume{46}.
\bdoi{10.1017/S0013091503214620}
\end{barticle}
\endbibitem

\bibitem{ADHD_DATA}
\begin{bmisc}[author]
\bauthor{\bsnm{Motie~Nasrabadi},~\bfnm{Ali}\binits{A.}},
  \bauthor{\bsnm{Allahverdy},~\bfnm{Armin}\binits{A.}},
  \bauthor{\bsnm{Samavati},~\bfnm{Mehdi}\binits{M.}} \AND
  \bauthor{\bsnm{Mohammadi},~\bfnm{Mohammad~Reza}\binits{M.~R.}}
(\byear{2020}).
\btitle{EEG data for ADHD / Control children}.
\bdoi{10.21227/rzfh-zn36}
\end{bmisc}
\endbibitem

\bibitem{USER_GUIDE_TDA}
\begin{barticle}[author]
\bauthor{\bsnm{Munch},~\bfnm{Elizabeth}\binits{E.}}
(\byear{2017}).
\btitle{A User’s Guide to Topological Data Analysis}.
\bjournal{Journal of Learning Analytics}
\bvolume{4}
\bpages{47–61}.
\bdoi{10.18608/jla.2017.42.6}
\end{barticle}
\endbibitem

\bibitem{MUNKRES_ALG_TOP}
\begin{bbook}[author]
\bauthor{\bsnm{Munkres},~\bfnm{James~R.}\binits{J.~R.}}
(\byear{1984}).
\btitle{Elements of Algebraic Topology}.
\bpublisher{Addison Wesley Publishing Company}.
\end{bbook}
\endbibitem

\bibitem{PERMUTATION_TESTING_NEUROIMAGING}
\begin{barticle}[author]
\bauthor{\bsnm{Nichols},~\bfnm{Thomas}\binits{T.}} \AND
  \bauthor{\bsnm{Holmes},~\bfnm{Andrew}\binits{A.}}
(\byear{2002}).
\btitle{Nonparametric permutation tests for functional neuroimaging: A primer
  with examples}.
\bjournal{Human brain mapping}
\bvolume{15}
\bpages{1-25}.
\bdoi{10.1002/hbm.1058}
\end{barticle}
\endbibitem

\bibitem{STAT_MODELS_EVO}
\begin{barticle}[author]
\bauthor{\bsnm{Ombao},~\bfnm{Hernando}\binits{H.}},
  \bauthor{\bsnm{Fiecas},~\bfnm{Mark}\binits{M.}},
  \bauthor{\bsnm{Ting},~\bfnm{Chee-Ming}\binits{C.-M.}} \AND
  \bauthor{\bsnm{Low},~\bfnm{Yin~Fen}\binits{Y.~F.}}
(\byear{2018}).
\btitle{Statistical models for brain signals with properties that evolve across
  trials}.
\bjournal{NeuroImage}
\bvolume{180}
\bpages{609-618}.
\bdoi{https://doi.org/10.1016/j.neuroimage.2017.11.061}
\end{barticle}
\endbibitem

\bibitem{SPECTRAL_DEPENDENCE}
\begin{bmisc}[author]
\bauthor{\bsnm{Ombao},~\bfnm{Hernando}\binits{H.}} \AND
  \bauthor{\bsnm{Pinto},~\bfnm{Marco}\binits{M.}}
(\byear{2021}).
\btitle{Spectral Dependence}.
\end{bmisc}
\endbibitem

\bibitem{ONBAO_BELLEGEM}
\begin{barticle}[author]
\bauthor{\bsnm{Ombao},~\bfnm{Hernando}\binits{H.}} \AND
  \bauthor{\bsnm{Van~Bellegem},~\bfnm{SÉbastien}\binits{S.}}
(\byear{2008}).
\btitle{Evolutionary Coherence of Nonstationary Signals}.
\bjournal{IEEE Transactions on Signal Processing}
\bvolume{56}
\bpages{2259-2266}.
\bdoi{10.1109/TSP.2007.914341}
\end{barticle}
\endbibitem

\bibitem{EXPERIMENTAL_RANDOMIZATION_1}
\begin{barticle}[author]
\bauthor{\bsnm{Raz},~\bfnm{Jonathan}\binits{J.}},
  \bauthor{\bsnm{Zheng},~\bfnm{Hui}\binits{H.}},
  \bauthor{\bsnm{Ombao},~\bfnm{Hernando}\binits{H.}} \AND
  \bauthor{\bsnm{Turetsky},~\bfnm{Bruce}\binits{B.}}
(\byear{2003}).
\btitle{Statistical tests for fMRI based on experimental randomization}.
\bjournal{NeuroImage}
\bvolume{19}
\bpages{226-232}.
\bdoi{https://doi.org/10.1016/S1053-8119(03)00115-0}
\end{barticle}
\endbibitem

\bibitem{TOPOLOGY_EULER}
\begin{bbook}[author]
\bauthor{\bsnm{Richeson},~\bfnm{David~S.}\binits{D.~S.}}
(\byear{2008}).
\btitle{Euler's Gem: The Polyhedron Formula and the Birth of Topology}.
\bpublisher{Princeton University Press}.
\end{bbook}
\endbibitem

\bibitem{HYPOTHESIS_TESTING_TDA}
\begin{barticle}[author]
\bauthor{\bsnm{Robinson},~\bfnm{Andrew}\binits{A.}} \AND
  \bauthor{\bsnm{Turner},~\bfnm{Katharine}\binits{K.}}
(\byear{2017}).
\btitle{Hypothesis Testing for Topological Data Analysis}.
\bjournal{Journal of Applied and Computational Topology}
\bvolume{1}
\bpages{241–261}.
\bdoi{10.1007/s41468-017-0008-7}
\end{barticle}
\endbibitem

\bibitem{TS_TDA}
\begin{binproceedings}[author]
\bauthor{\bsnm{Seversky},~\bfnm{Lee}\binits{L.}},
  \bauthor{\bsnm{Davis},~\bfnm{Shelby}\binits{S.}} \AND
  \bauthor{\bsnm{Berger},~\bfnm{Matthew}\binits{M.}}
(\byear{2016}).
\btitle{On Time-Series Topological Data Analysis: New Data and Opportunities}.
\bpages{1014-1022}.
\bdoi{10.1109/CVPRW.2016.131}
\end{binproceedings}
\endbibitem

\bibitem{TSA_SHUMWAY_STOFFER}
\begin{bbook}[author]
\bauthor{\bsnm{Shumway},~\bfnm{Robert~H.}\binits{R.~H.}} \AND
  \bauthor{\bsnm{Stoffer},~\bfnm{David~S.}\binits{D.~S.}}
(\byear{2005}).
\btitle{Time Series Analysis and Its Applications}.
\bpublisher{Springer-Verlag}.
\end{bbook}
\endbibitem

\bibitem{GRAPH_MODELING_COMPLEX_NETWORK}
\begin{barticle}[author]
\bauthor{\bsnm{Stam},~\bfnm{Cornelis}\binits{C.}} \AND
  \bauthor{\bsnm{Reijneveld},~\bfnm{Jaap}\binits{J.}}
(\byear{2007}).
\btitle{Graph theoretical analysis of complex networks in the brain}.
\bjournal{Nonlinear biomedical physics}
\bvolume{1}
\bpages{3}.
\bdoi{10.1186/1753-4631-1-3}
\end{barticle}
\endbibitem

\bibitem{TIME_DELAY_TAKENS}
\begin{barticle}[author]
\bauthor{\bsnm{Takens},~\bfnm{Floris}\binits{F.}}
(\byear{1981}).
\btitle{Detecting strange attractors in turbulence}.
\bjournal{Dynamical Systems and Turbulence, Lecture Notes in Mathematics}
\bvolume{898}
\bpages{366–381}.
\bdoi{http://crcv.ucf.edu/gauss/info/Takens.pdf}
\end{barticle}
\endbibitem

\bibitem{BIC_NET}
\begin{bmisc}[author]
\bauthor{\bsnm{Tang},~\bfnm{Meini}\binits{M.}},
  \bauthor{\bsnm{Ting},~\bfnm{Chee-Ming}\binits{C.-M.}} \AND
  \bauthor{\bsnm{Ombao},~\bfnm{Hernando}\binits{H.}}
(\byear{2021}).
\btitle{BICNet: A Bayesian Approach for Estimating Task Effects on Intrinsic
  Connectivity Networks in fMRI Data}.
\end{bmisc}
\endbibitem

\bibitem{DYN_CON_fMRI}
\begin{barticle}[author]
\bauthor{\bsnm{Ting},~\bfnm{Chee-Ming}\binits{C.-M.}},
  \bauthor{\bsnm{Ombao},~\bfnm{Hernando}\binits{H.}},
  \bauthor{\bsnm{Samdin},~\bfnm{S.~Balqis}\binits{S.~B.}} \AND
  \bauthor{\bsnm{Salleh},~\bfnm{Sh-Hussain}\binits{S.-H.}}
(\byear{2018}).
\btitle{Estimating Dynamic Connectivity States in fMRI Using Regime-Switching
  Factor Models}.
\bjournal{IEEE Transactions on Medical Imaging}
\bvolume{37}
\bpages{1011-1023}.
\bdoi{10.1109/TMI.2017.2780185}
\end{barticle}
\endbibitem

\bibitem{DYNAMIC_COMUNITY_STRUCTURE}
\begin{barticle}[author]
\bauthor{\bsnm{Ting},~\bfnm{Chee-Ming}\binits{C.-M.}},
  \bauthor{\bsnm{Samdin},~\bfnm{S.~Balqis}\binits{S.~B.}},
  \bauthor{\bsnm{Tang},~\bfnm{Meini}\binits{M.}} \AND
  \bauthor{\bsnm{Ombao},~\bfnm{Hernando}\binits{H.}}
(\byear{2021}).
\btitle{Detecting Dynamic Community Structure in Functional Brain Networks
  Across Individuals: A Multilayer Approach}.
\bjournal{IEEE Transactions on Medical Imaging}
\bvolume{40}
\bpages{468-480}.
\bdoi{10.1109/TMI.2020.3030047}
\end{barticle}
\endbibitem

\bibitem{TDA_BIOLOGY}
\begin{barticle}[author]
\bauthor{\bsnm{Topaz},~\bfnm{Chad~M.}\binits{C.~M.}},
  \bauthor{\bsnm{Ziegelmeier},~\bfnm{Lori}\binits{L.}} \AND
  \bauthor{\bsnm{Halverson},~\bfnm{Tom}\binits{T.}}
(\byear{2015}).
\btitle{Topological Data Analysis of Biological Aggregation Models}.
\bjournal{PLOS ONE}
\bvolume{10}
\bpages{1-26}.
\bdoi{10.1371/journal.pone.0126383}
\end{barticle}
\endbibitem

\bibitem{EEG_BRAIN_NETWORK}
\begin{barticle}[author]
\bauthor{\bsnm{{van Straaten}},~\bfnm{Elisabeth C.~W.}\binits{E.~C.~W.}} \AND
  \bauthor{\bsnm{Stam},~\bfnm{Cornelis~J.}\binits{C.~J.}}
(\byear{2013}).
\btitle{Structure out of chaos: Functional brain network analysis with EEG,
  MEG, and functional MRI}.
\bjournal{European Neuropsychopharmacology}
\bvolume{23}
\bpages{7-18}.
\bdoi{https://doi.org/10.1016/j.euroneuro.2012.10.010}
\end{barticle}
\endbibitem

\bibitem{FMRI_PAIN}
\begin{barticle}[author]
\bauthor{\bsnm{Wager},~\bfnm{Tor~D.}\binits{T.~D.}},
  \bauthor{\bsnm{Atlas},~\bfnm{Lauren~Y.}\binits{L.~Y.}},
  \bauthor{\bsnm{Lindquist},~\bfnm{Martin~A.}\binits{M.~A.}},
  \bauthor{\bsnm{Roy},~\bfnm{Mathieu}\binits{M.}},
  \bauthor{\bsnm{Woo},~\bfnm{Choong-Wan}\binits{C.-W.}} \AND
  \bauthor{\bsnm{Kross},~\bfnm{Ethan}\binits{E.}}
(\byear{2013}).
\btitle{An fMRI-Based Neurologic Signature of Physical Pain}.
\bjournal{New England Journal of Medicine}
\bvolume{368}
\bpages{1388-1397}.
\bdoi{10.1056/NEJMoa1204471}
\end{barticle}
\endbibitem

\bibitem{RELIABLE_FUNCTIONAL_CON}
\begin{barticle}[author]
\bauthor{\bsnm{Wang},~\bfnm{Yikai}\binits{Y.}},
  \bauthor{\bsnm{Kang},~\bfnm{Jian}\binits{J.}},
  \bauthor{\bsnm{Kemmer},~\bfnm{Phebe~B.}\binits{P.~B.}} \AND
  \bauthor{\bsnm{Guo},~\bfnm{Ying}\binits{Y.}}
(\byear{2016}).
\btitle{An Efficient and Reliable Statistical Method for Estimating Functional
  Connectivity in Large Scale Brain Networks Using Partial Correlation}.
\bjournal{Frontiers in Neuroscience}
\bvolume{10}.
\bdoi{10.3389/fnins.2016.00123}
\end{barticle}
\endbibitem

\bibitem{TDA_MORSE_EEG}
\begin{barticle}[author]
\bauthor{\bsnm{Wang},~\bfnm{Yuan}\binits{Y.}},
  \bauthor{\bsnm{Ombao},~\bfnm{Hernando}\binits{H.}} \AND
  \bauthor{\bsnm{Chung},~\bfnm{Moo~K.}\binits{M.~K.}}
(\byear{2018}).
\btitle{Topological Data Analysis of Single-Trial Electroencephalographic
  Signals}.
\bjournal{Annals of Applied Statistics}
\bvolume{12}
\bpages{1506–1534}.
\bdoi{10.1214/17-AOAS1119}
\end{barticle}
\endbibitem

\bibitem{SMALL_WORLD_NETWORKS}
\begin{barticle}[author]
\bauthor{\bsnm{Watts},~\bfnm{Duncan~J.}\binits{D.~J.}} \AND
  \bauthor{\bsnm{Strogatz},~\bfnm{Steven~H.}\binits{S.~H.}}
(\byear{1998}).
\btitle{Collective dynamics of ‘small-world’ networks}.
\bjournal{Nature}
\bvolume{393}
\bpages{440-442}.
\bdoi{10.1038/30918}
\end{barticle}
\endbibitem

\end{thebibliography}

%% or include bibliography directly:
% \begin{thebibliography}{}
% \bibitem{b1}
% \end{thebibliography}

\end{document}